\pdfoutput=1
\documentclass[12pt,a4paper]{article}
\usepackage{ifthen} % for conditional statements
\newboolean{pdflatex}
\setboolean{pdflatex}{true} % False for eps figures 

\newboolean{articletitles}
\setboolean{articletitles}{true} % False removes titles in references

\newboolean{uprightparticles}
\setboolean{uprightparticles}{false} %True for upright particle symbols

\newboolean{inbibliography}
\setboolean{inbibliography}{false} %True once you enter the bibliography

\usepackage{microtype}
\usepackage{lineno}  % for line numbering during review
\usepackage{xspace} % To avoid problems with missing or double spaces after
\usepackage{caption} %these three command get the figure and table captions automatically small

\usepackage{graphicx}  % to include figures (can also use other packages)
\usepackage{color}
\usepackage{colortbl}
\usepackage{amsmath} % Adds a large collection of math symbols
\usepackage{amssymb}
\usepackage{amsfonts}
\usepackage{upgreek} % Adds in support for greek letters in roman typeset
\usepackage{pstricks}
\usepackage{pst-plot}

\newcommand*\patchAmsMathEnvironmentForLineno[1]{%
\expandafter\let\csname old#1\expandafter\endcsname\csname #1\endcsname
\expandafter\let\csname oldend#1\expandafter\endcsname\csname
end#1\endcsname
 \renewenvironment{#1}%
   {\linenomath\csname old#1\endcsname}%
   {\csname oldend#1\endcsname\endlinenomath}%
}
\newcommand*\patchBothAmsMathEnvironmentsForLineno[1]{%
  \patchAmsMathEnvironmentForLineno{#1}%
  \patchAmsMathEnvironmentForLineno{#1*}%
}
\AtBeginDocument{%
\patchBothAmsMathEnvironmentsForLineno{equation}%
\patchBothAmsMathEnvironmentsForLineno{align}%
\patchBothAmsMathEnvironmentsForLineno{flalign}%
\patchBothAmsMathEnvironmentsForLineno{alignat}%
\patchBothAmsMathEnvironmentsForLineno{gather}%
\patchBothAmsMathEnvironmentsForLineno{multline}%
\patchBothAmsMathEnvironmentsForLineno{eqnarray}%
}

\usepackage{hyperref}    % Hyperlinks in references
\usepackage[all]{hypcap} % Internal hyperlinks to floats.

\usepackage{xspace} 
\usepackage{upgreek}

\def\lhcb {\mbox{LHCb}\xspace}

\def\MagUp {\mbox{\em Mag\kern -0.05em Up}\xspace}

\ifthenelse{\boolean{uprightparticles}}%
{
 
 \def\Pgamma      {\ensuremath{\upgamma}\xspace}

 \def\Pmu         {\ensuremath{\upmu}\xspace}

 \def\Ppi         {\ensuremath{\uppi}\xspace}

 \def\Ppsi        {\ensuremath{\uppsi}\xspace}

 \def\PDelta      {\ensuremath{\Delta}\xspace}                 
 \def\PXi      {\ensuremath{\Xi}\xspace}                 
 \def\PLambda      {\ensuremath{\Lambda}\xspace}                 
 \def\PSigma      {\ensuremath{\Sigma}\xspace}                 
 \def\POmega      {\ensuremath{\Omega}\xspace}                 
 \def\PUpsilon      {\ensuremath{\Upsilon}\xspace}                 
 
 \def\PB      {\ensuremath{\mathrm{B}}\xspace}                 
                  
 \def\PD      {\ensuremath{\mathrm{D}}\xspace}

 \def\PJ      {\ensuremath{\mathrm{J}}\xspace}                 
 \def\PK      {\ensuremath{\mathrm{K}}\xspace}

 \def\Pb      {\ensuremath{\mathrm{b}}\xspace}                 
 \def\Pc      {\ensuremath{\mathrm{c}}\xspace}

 \def\Pi      {\ensuremath{\mathrm{i}}\xspace}

 \def\Ps      {\ensuremath{\mathrm{s}}\xspace}

}
{
 
 \def\Pgamma      {\ensuremath{\gamma}\xspace}

 \def\Pmu         {\ensuremath{\mu}\xspace}

 \def\Ppi         {\ensuremath{\pi}\xspace}

 \def\Ppsi        {\ensuremath{\psi}\xspace}                 
                  
 \mathchardef\PDelta="7101
 \mathchardef\PXi="7104
 \mathchardef\PLambda="7103
 \mathchardef\PSigma="7106
 \mathchardef\POmega="710A
 \mathchardef\PUpsilon="7107
                  
 \def\PB      {\ensuremath{B}\xspace}                 
                  
 \def\PD      {\ensuremath{D}\xspace}

 \def\PJ      {\ensuremath{J}\xspace}                 
 \def\PK      {\ensuremath{K}\xspace}

 \def\Pb      {\ensuremath{b}\xspace}                 
 \def\Pc      {\ensuremath{c}\xspace}

 \def\Pi      {\ensuremath{i}\xspace}

 \def\Ps      {\ensuremath{s}\xspace}

}

\makeatletter
\ifcase \@ptsize \relax% 10pt
  \newcommand{\miniscule}{\@setfontsize\miniscule{4}{5}}% \tiny: 5/6
\or% 11pt
  \newcommand{\miniscule}{\@setfontsize\miniscule{5}{6}}% \tiny: 6/7
\or% 12pt
  \newcommand{\miniscule}{\@setfontsize\miniscule{5}{6}}% \tiny: 6/7
\fi
\makeatother

\DeclareRobustCommand{\optbar}[1]{\shortstack{{\miniscule (\rule[.5ex]{1.25em}{.18mm})}
  \\ [-.7ex] $#1$}}

\def\mup        {{\ensuremath{\Pmu^+}}\xspace}
\def\mun        {{\ensuremath{\Pmu^-}}\xspace} % muon negative (\mum is taken)

\def\g      {{\ensuremath{\Pgamma}}\xspace}

\def\squark    {{\ensuremath{\Ps}}\xspace}

\def\cquark    {{\ensuremath{\Pc}}\xspace}

\def\bquark    {{\ensuremath{\Pb}}\xspace}
\def\bquarkbar {{\ensuremath{\overline \bquark}}\xspace}
\def\bbbar     {{\ensuremath{\bquark\bquarkbar}}\xspace}

\def\pion   {{\ensuremath{\Ppi}}\xspace}

\def\pip    {{\ensuremath{\pion^+}}\xspace}
\def\pim    {{\ensuremath{\pion^-}}\xspace}

\def\kaon    {{\ensuremath{\PK}}\xspace}
  \def\Kbar    {{\kern 0.2em\overline{\kern -0.2em \PK}{}}\xspace}

\def\KorKbar    {\kern 0.18em\optbar{\kern -0.18em K}{}\xspace}

\def\Kp      {{\ensuremath{\kaon^+}}\xspace}
\def\Km      {{\ensuremath{\kaon^-}}\xspace}
\def\Kpm     {{\ensuremath{\kaon^\pm}}\xspace}

  \def\Dbar    {{\kern 0.2em\overline{\kern -0.2em \PD}{}}\xspace}
\def\D       {{\ensuremath{\PD}}\xspace}

\def\DorDbar    {\kern 0.18em\optbar{\kern -0.18em D}{}\xspace}
\def\Dz      {{\ensuremath{\D^0}}\xspace}
\def\Dzb     {{\ensuremath{\Dbar{}^0}}\xspace}

\def\Ds      {{\ensuremath{\D^+_\squark}}\xspace}
\def\Dsp     {{\ensuremath{\D^+_\squark}}\xspace}
\def\Dsm     {{\ensuremath{\D^-_\squark}}\xspace}

\def\B       {{\ensuremath{\PB}}\xspace}
\def\Bbar    {{\ensuremath{\kern 0.18em\overline{\kern -0.18em \PB}{}}}\xspace}
\def\Bb      {{\ensuremath{\Bbar}}\xspace}
\def\BorBbar    {\kern 0.18em\optbar{\kern -0.18em B}{}\xspace}

\def\Bu      {{\ensuremath{\B^+}}\xspace}
\def\Bub     {{\ensuremath{\B^-}}\xspace}
\def\Bp      {{\ensuremath{\Bu}}\xspace}

\def\Bpm     {{\ensuremath{\B^\pm}}\xspace}

\def\Bd      {{\ensuremath{\B^0}}\xspace}
\def\Bs      {{\ensuremath{\B^0_\squark}}\xspace}
\def\Bsb     {{\ensuremath{\Bbar{}^0_\squark}}\xspace}

\def\jpsi     {{\ensuremath{{\PJ\mskip -3mu/\mskip -2mu\Ppsi\mskip 2mu}}}\xspace}

  \def\Y#1S{\ensuremath{\PUpsilon{(#1S)}}\xspace}% no space before {...}!

\def\Lz          {{\ensuremath{\PLambda}}\xspace}
\def\Lbar        {{\ensuremath{\kern 0.1em\overline{\kern -0.1em\PLambda}}}\xspace}
\def\LorLbar    {\kern 0.18em\optbar{\kern -0.18em \PLambda}{}\xspace}

\def\Lb      {{\ensuremath{\Lz^0_\bquark}}\xspace}

\newcommand{\decay}[2]{\ensuremath{#1\!\to #2}\xspace}         % {\Pa}{\Pb \Pc}

\def\to                 {\ensuremath{\rightarrow}\xspace}

\def\CP                {{\ensuremath{C\!P}}\xspace}

\newcommand{\dms}{{\ensuremath{\Delta m_{\squark}}}\xspace}

\newcommand{\DGs}{{\ensuremath{\Delta\Gamma_{\squark}}}\xspace}

\newcommand{\phis}{{\ensuremath{\phi_{\squark}}}\xspace}

\newcommand{\etag}{{\ensuremath{\varepsilon_{\mathrm{tag}}}}\xspace}

\newcommand{\effeff}{\ensuremath{\varepsilon_{\mathrm{eff}}}\xspace}

\def\AT#1     {\ensuremath{A_{\mathrm{T}}^{#1}}\xspace}           % 2

\def\C#1      {\ensuremath{\mathcal{C}_{#1}}\xspace}                       % 9
\def\Cp#1     {\ensuremath{\mathcal{C}_{#1}^{'}}\xspace}                    % 7
\def\Ceff#1   {\ensuremath{\mathcal{C}_{#1}^{\mathrm{(eff)}}}\xspace}        % 9  
\def\Cpeff#1  {\ensuremath{\mathcal{C}_{#1}^{'\mathrm{(eff)}}}\xspace}       % 7
\def\Ope#1    {\ensuremath{\mathcal{O}_{#1}}\xspace}                       % 2
\def\Opep#1   {\ensuremath{\mathcal{O}_{#1}^{'}}\xspace}                    % 7

\newcommand{\tev}{\ifthenelse{\boolean{inbibliography}}{\ensuremath{~T\kern -0.05em eV}\xspace}{\ensuremath{\mathrm{\,Te\kern -0.1em V}}}\xspace}
\newcommand{\gev}{\ensuremath{\mathrm{\,Ge\kern -0.1em V}}\xspace}
\newcommand{\mev}{\ensuremath{\mathrm{\,Me\kern -0.1em V}}\xspace}
\newcommand{\kev}{\ensuremath{\mathrm{\,ke\kern -0.1em V}}\xspace}
\newcommand{\ev}{\ensuremath{\mathrm{\,e\kern -0.1em V}}\xspace}
\newcommand{\gevc}{\ensuremath{{\mathrm{\,Ge\kern -0.1em V\!/}c}}\xspace}
\newcommand{\mevc}{\ensuremath{{\mathrm{\,Me\kern -0.1em V\!/}c}}\xspace}
\newcommand{\gevcc}{\ensuremath{{\mathrm{\,Ge\kern -0.1em V\!/}c^2}}\xspace}
\newcommand{\gevgevcccc}{\ensuremath{{\mathrm{\,Ge\kern -0.1em V^2\!/}c^4}}\xspace}
\newcommand{\mevcc}{\ensuremath{{\mathrm{\,Me\kern -0.1em V\!/}c^2}}\xspace}

\def\mum  {\ensuremath{{\,\upmu\mathrm{m}}}\xspace}

\def\invfb   {\ensuremath{\mbox{\,fb}^{-1}}\xspace}

\newcommand{\stat}{\ensuremath{\mathrm{\,(stat)}}\xspace}
\newcommand{\syst}{\ensuremath{\mathrm{\,(syst)}}\xspace}

\newcommand{\chisq}{\ensuremath{\chi^2}\xspace}
\newcommand{\chisqndf}{\ensuremath{\chi^2/\mathrm{ndf}}\xspace}

\def\gsim{{~\raise.15em\hbox{$>$}\kern-.85em
          \lower.35em\hbox{$\sim$}~}\xspace}
\def\lsim{{~\raise.15em\hbox{$<$}\kern-.85em
          \lower.35em\hbox{$\sim$}~}\xspace}

\def\sWeights{\mbox{\em sWeights}\xspace}

\def\ptot       {\mbox{$p$}\xspace}
\def\pt         {\mbox{$p_{\mathrm{ T}}$}\xspace}

\def\mrad{\ensuremath{\mathrm{ \,mrad}}\xspace}
\def\rad{\ensuremath{\mathrm{ \,rad}}\xspace}

\def\evtgen     {\mbox{\textsc{EvtGen}}\xspace}

\def\geant      {\mbox{\textsc{Geant4}}\xspace}

\def\photos     {\mbox{\textsc{Photos}}\xspace}

\def\pythia     {\mbox{\textsc{Pythia}}\xspace}

\def\tell1  {TELL1\xspace}
\def\ukl1   {UKL1\xspace}

\newcommand{\eg}{\mbox{\itshape e.g.}\xspace}
\newcommand{\ie}{\mbox{\itshape i.e.}\xspace}

\def\Bstwo    {{\ensuremath{\B_{\squark2}^{*}(5840)^0}}\xspace}
\def\BstwoBK   {\mbox{\decay{\Bstwo}{\Bu\Km}}\xspace}

\def\BsDspi   {\mbox{\decay{\Bs}{\Dsm\pip}}\xspace} 
 
\usepackage{cite} % Allows for ranges in citations
\usepackage{mciteplus}

\begin{document}

%%%%%%%%%%%%%%%%%%%%%%%%%
%%%%% Title     %%%%%%%%%
%%%%%%%%%%%%%%%%%%%%%%%%%
\renewcommand{\thefootnote}{\fnsymbol{footnote}}
\setcounter{footnote}{1}

% $Id: title-LHCb-PAPER.tex 83530 2015-11-03 21:01:14Z smenzeme $
% ===============================================================================
% Purpose: LHCb-PAPER journal paper title page template
% Author: 
% Created on: 2010-09-25
% ===============================================================================

%%%%%%%%%%%%%%%%%%%%%%%%%
%%%%%  TITLE PAGE  %%%%%%
%%%%%%%%%%%%%%%%%%%%%%%%%
\begin{titlepage}
\pagenumbering{roman}

% Header ---------------------------------------------------
\vspace*{-1.5cm}
\centerline{\large EUROPEAN ORGANIZATION FOR NUCLEAR RESEARCH (CERN)}
\vspace*{1.5cm}
\noindent
\begin{tabular*}{\linewidth}{lc@{\extracolsep{\fill}}r@{\extracolsep{0pt}}}
\ifthenelse{\boolean{pdflatex}}% Logo format choice
{\vspace*{-2.7cm}\mbox{\!\!\!\includegraphics[width=.14\textwidth]{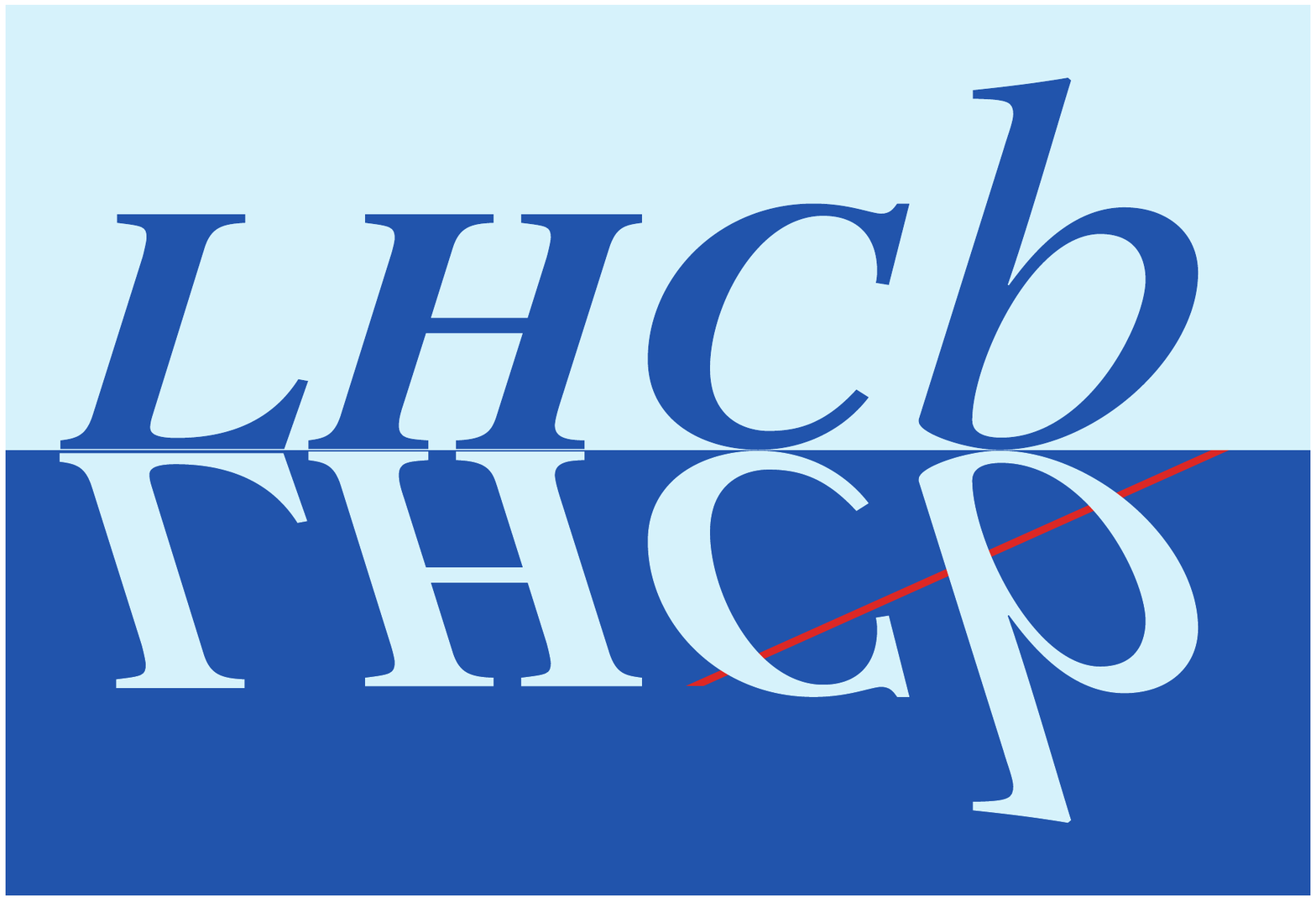}} & &}%
{\vspace*{-1.2cm}\mbox{\!\!\!\includegraphics[width=.12\textwidth]{lhcb-logo.eps}} & &}%
\\
 & & CERN-EP-2016-030 \\  % ID 
 & & LHCb-PAPER-2015-056 \\  % ID 
 & & February 23, 2016 \\ % \today
 & & \\
% not in paper \hline
\end{tabular*}

\vspace*{2.0cm}

% Title --------------------------------------------------
{\bf\boldmath\huge
\begin{center}
  A new algorithm for identifying \\ 
  the flavour of \Bs mesons at LHCb \\ 
\end{center}
}
% Authors -------------------------------------------------
\begin{center}
The LHCb collaboration\footnote{Authors are listed at the end of this paper.}
\end{center}
\vspace*{1.0cm}

% Abstract -----------------------------------------------
\begin{abstract}
  \noindent
  A new algorithm for the determination of the initial flavour of \Bs mesons is presented.  
 The algorithm is based on two neural networks and exploits the \bquark~hadron 
production mechanism at a hadron collider. 
The first network is trained to select charged kaons produced in
 association with the \Bs  meson. The second network combines the kaon charges to
assign the \Bs flavour and estimates the probability of a wrong assignment. 
The algorithm is calibrated using
data corresponding to an integrated luminosity of 3\invfb collected by the LHCb
experiment in proton-proton collisions at 7 and 8\tev centre-of-mass energies. 
The calibration is performed in two ways: by resolving the $\Bs$-$\Bsb$
flavour oscillations  
in \BsDspi decays, and by analysing
flavour-specific \BstwoBK decays. 
The tagging power measured in \BsDspi decays is
found to be  $(1.80 \pm 0.19\stat \pm 0.18\syst)$\%, 
which is an improvement of about 50\% compared to 
a similar algorithm previously used 
in the LHCb experiment.  
\end{abstract}

\vspace*{1.0cm}

\begin{center}
Published in JINST 11 (2016) P05010
\end{center}

\vspace{\fill}

{\footnotesize 
\centerline{\copyright~CERN on behalf of the \lhcb collaboration, licence \href{http://creativecommons.org/licenses/by/4.0/}{CC-BY-4.0}.}}
\vspace*{2mm}

\end{titlepage}

%%%%%%%%%%%%%%%%%%%%%%%%%%%%%%%%
%%%%%  EOD OF TITLE PAGE  %%%%%%
%%%%%%%%%%%%%%%%%%%%%%%%%%%%%%%%

%  empty page follows the title page ----
\newpage
\setcounter{page}{2}
\mbox{~}

\cleardoublepage

%\twocolumn
% %%%%%%%%%%%%% ---------

\renewcommand{\thefootnote}{\arabic{footnote}}
\setcounter{footnote}{0}

%%%%%%%%%%%%%%%%%%%%%%%%%
%%%%% Main text %%%%%%%%%
%%%%%%%%%%%%%%%%%%%%%%%%%

\pagestyle{plain} % restore page numbers for the main text
\setcounter{page}{1}
\pagenumbering{arabic}

\section{Introduction}
\label{sec:Introduction}

Precision measurements of flavour oscillations of $B^0_{(s)}$ mesons  and 
of \CP asymmetries in their decays allow 
the validity of the standard model of particle physics  
to be probed at energy scales not directly accessible by current colliders~\cite{LHCb-PAPER-2012-031}.  
Measurements of associated observables, \eg the  \CP-violating phase \phis  
in $\mbox{\decay{\Bs}{\jpsi K^+K^-}}$ and $\mbox{\decay{\Bs}{\jpsi \pi^+\pi^-}}$  
decays~\cite{LHCb-PAPER-2014-059,LHCb-PAPER-2014-019}, 
are among the major goals of the LHCb experiment and its upgrade~\cite{LHCb_roadmap,LHCb_upgrade}.\footnote{The inclusion of charge-conjugate 
decays is implied throughout this paper unless otherwise stated.}  
These analyses require so-called flavour-tagging algorithms  
to identify the flavour at production of the reconstructed \B meson. 
Improving the effectiveness of those algorithms is of crucial importance, 
as it increases the statistical power of the dataset collected by an experiment.

Several types of flavour-tagging algorithms have been developed in 
experiments at hadron colliders.  
Opposite-side (OS) algorithms exploit the fact that \bquark quarks 
are predominantly produced in \bbbar pairs in hadron collisions, 
and thus the flavour at production of the reconstructed \B meson is opposite 
to that of the other \bquark hadron in the event.  
Therefore, the products of the decay chain of the other \bquark
 hadron can be used for flavour tagging. 
The OS algorithms utilised in LHCb are described in 
Refs.~\cite{LHCb-PAPER-2011-027,LHCb-PAPER-2015-027}. 
Same-side (SS) algorithms look for particles produced in association 
with the reconstructed \B meson in the hadronisation process~\cite{Gronau:1992ke,Abe:1998sq,Abulencia:2006mq}. 
In about 50\% of cases, a \Bs meson is accompanied by a charged 
kaon and a \Bd meson by a charged pion. 
The charge of these particles indicates the \bquark quark content of the \B meson.
Information from OS and SS algorithms is usually combined in flavour-tagged analyses.  

This paper describes a new same-side kaon (SSK) flavour-tagging algorithm  
 at the LHCb experiment.  The first use of an SSK algorithm in LHCb is reported in
Refs.~\cite{LHCb-PAPER-2013-002,LHCb-PAPER-2013-006}. 
That version uses a selection algorithm, optimised with data, to identify the kaons
produced in the hadronisation of the \Bs meson. 
One key part of the algorithm is that, for events in which 
several particles pass the selection, the one with 
the largest transverse
momentum is chosen as the tagging candidate and its charge defines the tagging decision. 
The new algorithm presented here 
exploits two neural networks to identify the flavour at production of a reconstructed \Bs meson. 
The first neural network is used to assign  to each  
track reconstructed in the $pp$ collision a probability of being a particle related to the \Bs
hadronisation process. 
Tracks that have a probability larger than a suitably chosen threshold are combined in the  
second neural network to determine the tagging decision.

The effectiveness of an  
algorithm to tag a sample of reconstructed \B candidates 
is quantified by the tagging efficiency, $\etag$, and the mistag fraction, $\omega$. 
These variables are defined as
\begin{equation}
\etag = \frac{R+W}{R+W+U},\text{ and } \omega = \frac{W}{R+W},
 \label{tagging_variable}
\end{equation}
where $R$, $W$ and $U$ are the number of correctly tagged, incorrectly tagged,
and untagged \B candidates, respectively.  
For each tagged \B candidate $i$, the flavour-tagging algorithm estimates the probability, $\eta_i$, 
of an incorrect tag decision. 
%The mistag probability is defined such that, 
%  considering the set of tagged candidates 
% in any interval of the $\eta$ distribution, 
% the average of the $\eta_i$ values is the mistag fraction $\omega$ of that set.  
To correct for potential biases in $\eta_i$,  
 a function $\omega(\eta)$ is used to calibrate the mistag probability 
 to provide an unbiased estimate of the mistag fraction for any value of $\eta$.  
 The tagging efficiency and mistag probabilities are used to calculate
the effective tagging efficiency, $\effeff$, also known as the tagging power, 
  \begin{equation}
  \label{eq:effeff}
\effeff  = \etag  \frac{1}{R+W}
\sum_{i=1}^{R+W}\left (1-2\omega(\eta_i)\right)^2, 
\end{equation}
which represents the figure of merit in the optimisation 
of a flavour-tagging algorithm, since  
 the overall statistical power of the flavour-tagged sample is proportional to $\effeff$.  
 The previous SSK algorithm used by the LHCb experiment 
 has a tagging power of 0.9\% and 1.2\%
 in $\decay{\Bs}{\jpsi\phi}$ and \BsDspi decays, respectively. 
 For comparison, the performance of the combination of the OS algorithms in these 
 decays corresponds to a tagging power of
 about 2.3\% and 2.6\%~\cite{LHCb-PAPER-2013-002,LHCb-PAPER-2013-006}.

 The calibration function $\omega(\eta)$ is obtained with 
control samples of flavour-specific decays, \ie decays in which the \B flavour at decay  
 is known from the charge of the final-state particles.  
In the case of the new SSK algorithm described here, 
the decay \BsDspi\ and, for the first time, the decay \BstwoBK are used. 
These decays are reconstructed in a dataset corresponding 
to an integrated luminosity of 3\invfb collected by LHCb
in $pp$ collisions at 7 and 8\tev centre-of-mass energies.

\section{Detector and simulation}
\label{sec:Detector}
The \lhcb detector~\cite{Alves:2008zz,LHCb-DP-2014-002} is a single-arm forward
spectrometer covering the \mbox{pseudorapidity} range between 2 and 5,
designed for the study of particles containing \bquark or \cquark
quarks. The detector includes a high-precision tracking system
consisting of a silicon-strip vertex detector surrounding the $pp$
interaction region, a large-area silicon-strip detector located
upstream of a dipole magnet with a bending power of about
$4{\rm\,Tm}$, and three stations of silicon-strip detectors and straw
drift tubes placed downstream of the magnet. 
The polarity of the dipole magnet is reversed periodically throughout data-taking
to reduce the effect of asymmetries in the detection of charged particles.
The tracking system provides a measurement of momentum, \ptot, of charged particles with
a relative uncertainty that varies from 0.5\% at low momentum to 1.0\% at 200\gevc.
The minimum distance of a track to a primary $pp$ interaction vertex (PV), the impact parameter, is
measured with a resolution of $(15+29/\pt)\mum$,
where \pt is the component of the momentum transverse to the beam, in\gevc.
Different types of charged hadrons are distinguished using information
from two ring-imaging Cherenkov detectors. 
Photons, electrons and hadrons are identified by a calorimeter system consisting of
scintillating-pad and preshower detectors, an electromagnetic
calorimeter and a hadronic calorimeter. Muons are identified by a
system composed of alternating layers of iron and multiwire
proportional chambers. 
The online event selection is performed by a trigger~\cite{LHCb-DP-2012-004},
 which consists of a hardware stage and a software stage. 
 At the hardware trigger stage, for decay candidates of interest in this paper, 
 events are required to have a
 hadron with high transverse energy in the calorimeters, 
 or muons with high \pt. For hadrons, the transverse energy threshold is 3.5\gev.
% For muons, a transverse momentum $\pt>1.48\gevc$
% in the 7\tev data or $\pt>1.76\gevc$ in the 8\tev data is required.
% In case of B decays with a $\Jpsi \to \mup\mum$ in the final state,  
% the two muons must have ${\sqrt{\pt(\mu_1)\,\pt(\mu_2)}>1.3\gevc}$ 
% in 7\tev data or ${\sqrt{\pt(\mu_1)\,\pt(\mu_2)}>1.6\gevc}$ in the 8\tev data.  
 The software trigger requires a two-, three- or four-track
 secondary vertex with a significant displacement from the primary
  vertices. At least one charged particle
 must have a transverse momentum $\pt > 1.7\gevc$ and be
 inconsistent with originating from a PV.
 A multivariate algorithm~\cite{BBDT} is used for
 the identification of secondary vertices consistent with the decay
 of a \bquark hadron.
 
In the simulation, $pp$ collisions are generated using
\pythia~\cite{Sjostrand:2006za,*Sjostrand:2007gs} 
 with a specific \lhcb
configuration~\cite{LHCb-PROC-2010-056}.  Decays of hadronic particles
are described by \evtgen~\cite{Lange:2001uf}, in which final-state
radiation is generated using \photos~\cite{Golonka:2005pn}. The
interaction of the generated particles with the detector, and its response,
are implemented using the \geant
toolkit~\cite{Allison:2006ve, *Agostinelli:2002hh} as described in
Ref.~\cite{LHCb-PROC-2011-006}.

\section{The neural-network-based SSK algorithm }
\label{sec:strategy}

In this section, charged kaons related to the fragmentation process of the reconstructed \Bs candidate are called
signal, and other particles in the event are called background. This background includes, for
example, the decay products  of the 
OS \bquark hadron, and particles originating from soft QCD processes in $pp$ interactions.  
In the neural-network-based SSK algorithm, a neural network (NN1) 
classifies as signal or background
all tracks  passing an initial preselection.
A second neural network (NN2) combines 
the tracks selected by NN1 to 
tag the reconstructed $B$ candidate as either \Bs or \Bsb, and estimates 
the mistag probability associated with the tagging decision. 
Both NN1 and NN2 are based on the algorithms of Ref.~\cite{BDT}.

The preselection imposes a number of requirements on the tracks 
to be considered as tagging candidates, and is common to other flavour-tagging 
algorithms used in LHCb~\cite{LHCb-PAPER-2011-027}.
The tracks must have been measured in at least one of the tracking stations both
before and after the magnet.  Their momentum is required 
to be larger than 2\gevc, and their transverse momentum to be smaller than 10\gevc. 
A requirement that the angle between the tracks 
and the beam line must be at least 12\mrad is applied, 
 to reject particles which  either originate from interactions with the beam pipe material 
 or which suffer from multiple scattering in this region. 
 The tracks associated with the reconstructed decay products of the \Bs candidate are excluded. 
 Tracks in a cone of 5\mrad around the \Bs flight direction are rejected 
   to remove any remaining \Bs decay products. 
  Tracks outside a cone of 1.5\rad are also rejected, to suppress particles which are
   not correlated with the \Bs flavour. 
Finally, tracks must be inconsistent  
with originating at a different PV  
from the one associated with the reconstructed \Bs candidate, 
which is taken to be that closest to the \Bs flight path.

The network NN1 is trained using signal and background kaons from approximately 80,000 simulated events 
containing a reconstructed $\decay{\Bs}{\decay{D_s^-(}{K^+K^-\pi^-)}\pi^+}$ decay. 
An independent sample of similar size 
is used to test the network's performance. 
Information from the simulation is used to ensure that only genuine, 
correctly reconstructed \BsDspi decays are used. 
The following ten variables are used as input to NN1: 
the momentum and transverse momentum of the track;
 the \chisq per degree of freedom of the track fit;
 the track impact parameter significance, defined as 
 the ratio between the track impact parameter with respect to the 
 PV associated with the \Bs candidate, and its uncertainty;
the difference of the transverse momenta of the track and the \Bs candidate; 
 the difference of the azimuthal angles and of the pseudorapidities between the track and the \Bs candidate;
 the number of reconstructed primary vertices; the number of tracks passing the preselection; and the 
transverse momentum of the \Bs candidate.  
 The track impact parameter significance is used to quantify the 
 probability that a track originates from the same primary vertex as the reconstructed \Bs candidate. 
In an event with a large number of tracks and primary vertices,
the probability that a given track is a signal fragmentation track is lower;  
hence the use of these variables in  NN1. 
The \Bs transverse momentum is correlated with the difference 
in pseudorapidity of the fragmentation
tracks and the \Bs candidate. 

\begin{figure}[t]
\centering
\includegraphics[width=.45\columnwidth]{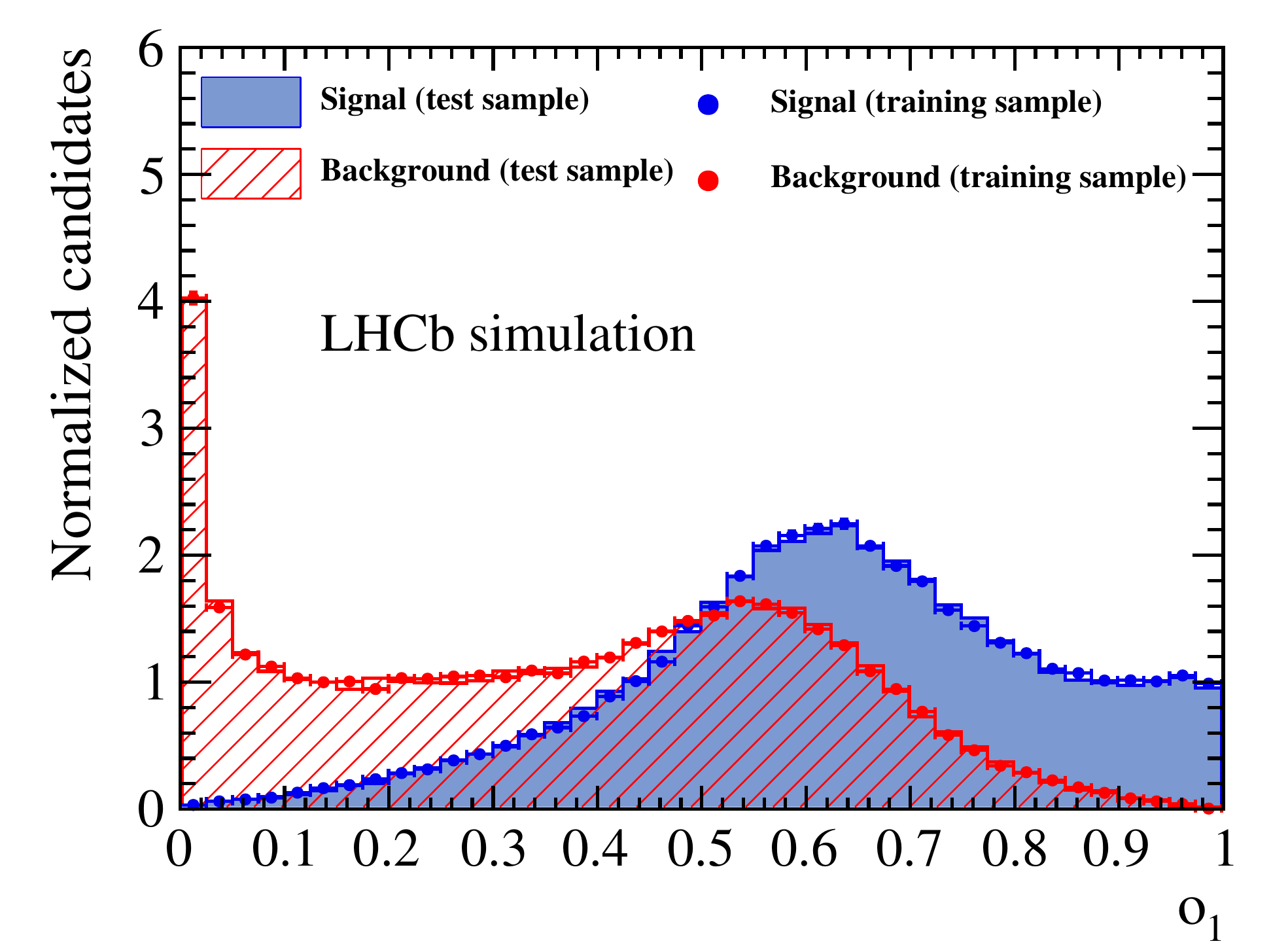}~~~
\includegraphics[width=.45\columnwidth]{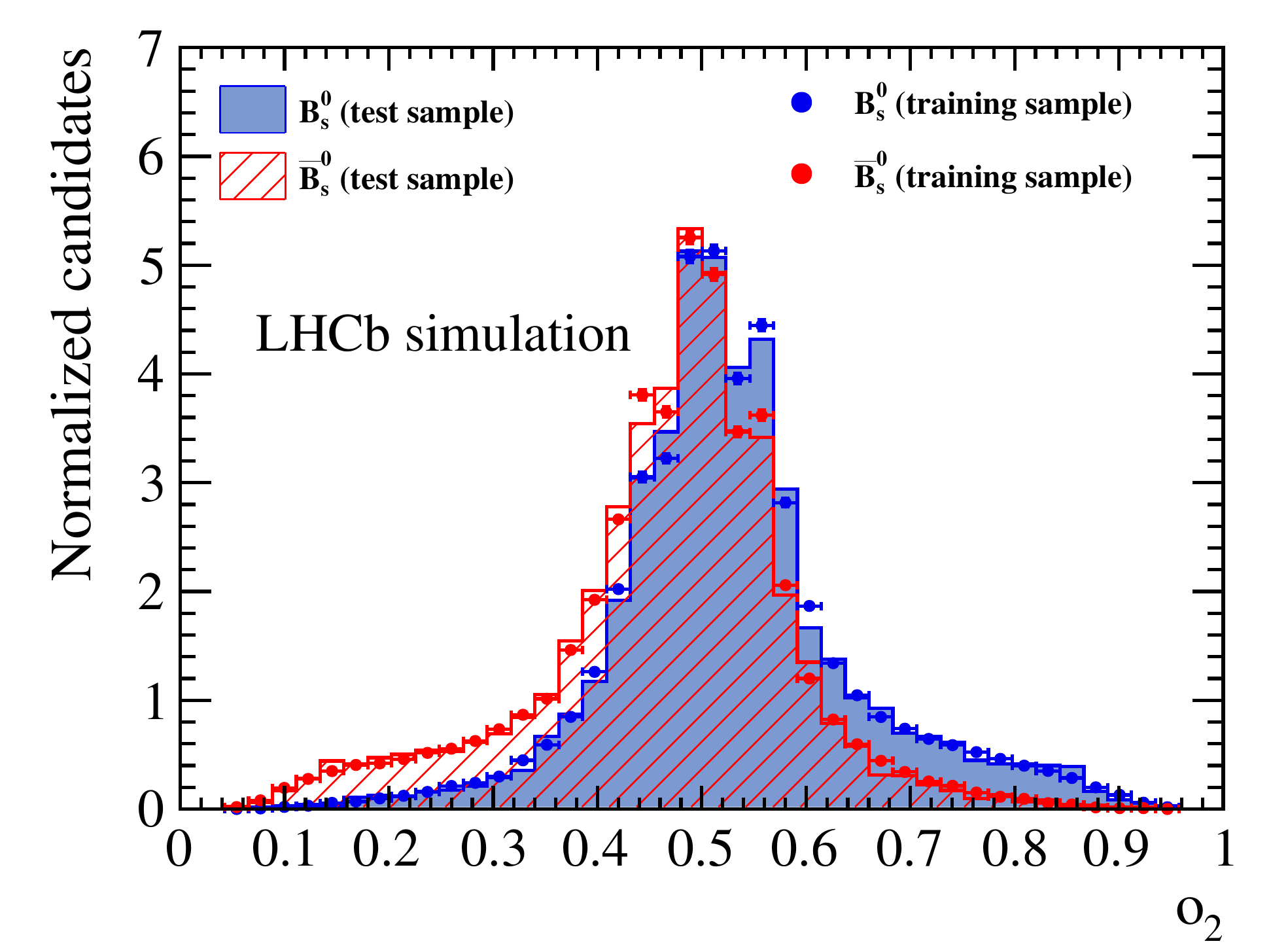}
\caption{(left)~Distribution of the NN1 output, $o_1$, of signal (blue) and background (red) tracks.
(right)~Distribution of the NN2 output, $o_2$, of initially produced \Bs (blue) and \Bsb (red) mesons. 
Both distributions are obtained with simulated events. 
The markers represent the distributions obtained from the training samples; 
the solid histograms are the distributions obtained from the test samples. 
The good agreement between the distributions of the test and training samples shows 
that there is no overtraining of the classifiers. 
\label{fig:NN2_prob}}
\end{figure}

The network NN1 features one hidden layer with nine nodes. 
The activation function and the estimator type are chosen following 
the recommendations of Ref.~\cite{Breiman}, to guarantee the 
probabilistic interpretation of the response function.
The distribution of the NN1 output, $o_1$, for signal and background candidates
 is illustrated in Fig.~\ref{fig:NN2_prob}. 
 After requiring $o_1 > 0.65$, 
about 60\% of the reconstructed \BsDspi decays have at least one tagging candidate in background-subtracted data.
This number corresponds to the tagging efficiency. 
 The network configuration and the $o_1$ requirement are chosen to give the largest tagging power. 
For each tagged \Bs candidate there are on average 1.6 tagging tracks, 
to be combined in NN2.  

The training of NN2 is carried out with a  
simulated sample of approximately 80,000 
reconstructed \BsDspi decays,  
 statistically independent of that used to train NN1. 
 All of the events contain at least one track
  passing the NN1 selection requirement. 
  Half of the events contain a meson whose 
  true initial flavour is \Bs, and the other half contain \Bsb mesons. 
About 90\% of the simulated events are used
to train NN2, and the remaining 10\% are used to test its performance. 
The likelihood of the track of being a kaon~\cite{LHCb-DP-2014-002} and 
the value of $o_1$ are used as input variables to NN2. 
These variables are multiplied by the charge of the tagging track,
to exploit the charge correlation of fragmentation kaons
with the flavour of the \Bs meson.
The reconstructed \Bs momentum, its transverse momentum,
the number of reconstructed primary vertices and the number of
reconstructed tracks in the event that pass the \Bs candidate's selection  
are also used as input to NN2. 
Different configurations of NN2 with up to $n_{\rm max}$ input tagging tracks and several network
structures are tested. In all cases, one hidden layer with $n-1$ nodes
is chosen, where $n$ is the number of input variables. 
If more than $n_{\rm max}$ tracks pass the requirement on $o_1$, 
the $n_{\rm max}$ tracks with the greatest $o_1$ are used. 
If fewer than $n_{\rm max}$ pass, the unused input values are set to zero.
The networks with $n_{\rm max}=2$, 3 and 4 perform very similarly and show a significantly
better separation than the configurations with $n_{\rm max}=1$ or 5. The NN2
configuration with $n_{\rm max}=3$ is chosen. 
%The main additional tagging power of this algorithm with respect to the 
%previous SSK algorithm comes from the treatment of 
%events with two tagging tracks of very similar quality.
The main additional tagging power of this algorithm compared 
to the previous SSK algorithm comes from the possibility to treat events 
with multiple tracks of similar tagging quality, 
which allows a looser selection (\ie a larger tagging efficiency) 
compared to the algorithm using a single tagging track.
The distribution of the NN2 output, $o_2$, of initially produced \Bs and \Bsb mesons 
is shown in Fig.~\ref{fig:NN2_prob}.

In the training configuration used~\cite{Breiman}, the NN2 output 
 can be directly interpreted as the probability that a \B candidate with a given value of $o_2$ 
 was initially produced as a  \Bs meson, 
\begin{equation}
P(\Bs|o_2)= o_2 = \frac{N_{\Bs}(o_2)}{N_{\Bs}(o_2) +
  N_{\Bsb}(o_2)},
\end{equation}
where the second equality holds in the limit of infinite statistics, and 
$N_{\Bs}(o_2)$ and $N_{\Bsb}(o_2)$ refer to the number of
initial \Bs and \Bsb mesons in the training
sample with a given $o_2$ value. The distribution of the NN2 output of initial \Bs mesons has
a peak at $o_2$ values slightly larger than 0.5, while that of
initial \Bsb mesons has a peak at $o_2$ values slightly smaller than
0.5 (Fig.~\ref{fig:NN2_prob}). In case of no \CP 
asymmetries, and no asymmetries related to the different interaction probabilities 
of charged kaons with the detector, 
the NN2 distribution of initial \Bs mesons is expected to be identical, within uncertainties, to the NN2 
distribution of initial \Bsb mesons mirrored at $o_2=0.5$. 
This is a prerequisite for interpreting the NN2 output as a mistag probability. Therefore, 
to ensure such an interpretation, a new variable is defined,
which has a mirrored distribution for initial \Bs and \Bsb mesons of the same kinematics, 
\begin{equation}
o_2' = \frac{o_2 + (1- \bar{o}_2)}{2} \label{eq:nn2prime},
\end{equation} 
where $\bar{o}_2$ stands for the NN2 output with the charged-conjugated input variables, \ie   
for a specific candidate, $\bar{o}_2$ is 
evaluated by flipping the charge signs of the input variables of NN2.
The tagging decision is defined such that the \B candidate is assumed 
to be produced as a \Bs if $o_2' > 0.5$ and as a \Bsb if $o_2' < 0.5$.
Likewise, the mistag probability is defined as
$\eta = 1- o_2'$ for candidates tagged as \Bs,  and as 
$\eta = o_2'$ for candidates tagged as \Bsb.

\section{Calibration using {\bf \boldmath $B^0_s \rightarrow D^-_s \pi^+$} decays }
\label{sec:cal}

The mistag probability estimated by the SSK algorithm is calibrated
using two different decays, \BsDspi and  \BstwoBK. The calibration with \BsDspi
decays requires the \Bs-\Bsb flavour oscillations  to be resolved  
via a fit to the \Bs decay time distribution,  since  the amplitude of
the oscillation is related to the mistag fraction. 
In contrast, there are no flavour oscillations before the strong decay of the \Bstwo and the
charged mesons produced in its decays directly identify the \Bstwo
production flavour. Therefore, the calibration with \Bstwo 
is performed by counting the number of correctly and incorrectly
tagged signal candidates. Thus, the two calibrations feature different analysis
 techniques, which are affected by different 
sources of systematic uncertainties, and serve as cross-checks of each other. 
The calibration with \BsDspi decays is described in this section and that using
\BstwoBK decays in Sect.~\ref{Bsstar}. 
The results are combined in Sect.~\ref{sec:calibration_summary} after 
equalising the transverse momentum spectra of the reconstructed \Bs
and \Bstwo candidates,  since the calibration parameters depend on 
 the kinematics of the reconstructed \B decay. 
These calibrations also serve as a test of the new  algorithm
in data, to evaluate the performance of the tagger and 
to compare it to that of the previous SSK algorithm used in LHCb.

\begin{figure}[tb]
 \centering
\includegraphics[width=.55\textwidth]{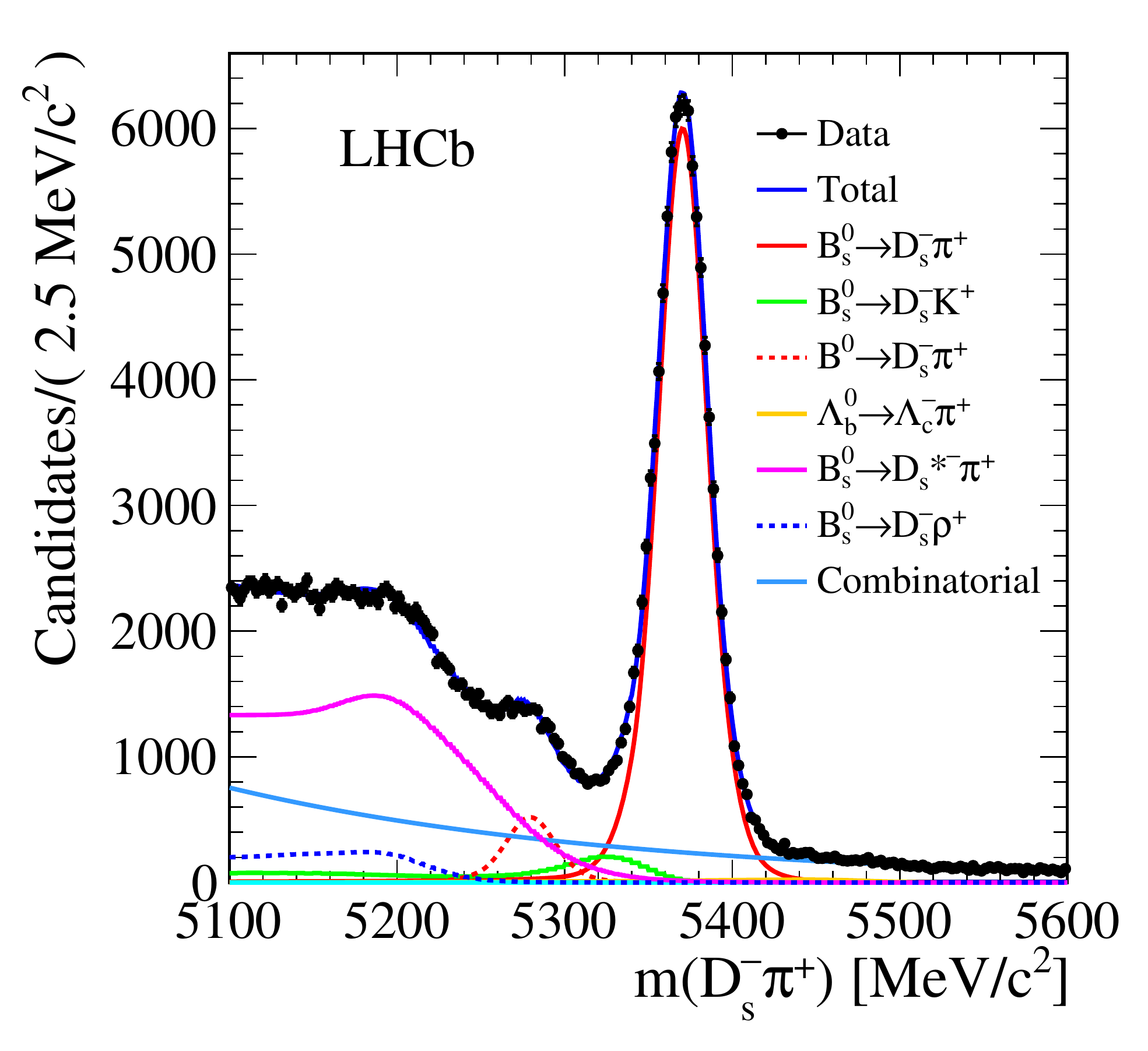}
\caption{Mass distribution of \BsDspi candidates with fit
projections overlaid.  
Data points (black markers) correspond to the \Bs candidates selected in the 3\invfb data sample. 
The total fit function and its components are overlaid with solid and dashed lines (see legend). 
}\label{fig:mass}
\end{figure}

A sample of \BsDspi candidates is selected
according to the requirements presented in Ref.~\cite{LHCb-PAPER-2014-038}. The
\Dsm candidates are reconstructed in the final states $\Kp\Km\pim$ and  $\pim\pip\pim$.
 The $\Dsm \pip$ mass spectrum contains a
narrow peak, corresponding to \BsDspi signal candidates, 
and other broader structures due to misreconstructed \bquark-hadron decays, all on top
of a smooth background distribution due to random combinations of tracks
passing the selection requirements. The
 signal and background components are determined by a fit to the mass 
 distribution of candidates in the range 
5100--5600\mevcc (Fig.~\ref{fig:mass}). 
The signal component is described as the sum 
of two Gaussian functions with a common mean, 
plus a power-law tail on each side, which is fixed from simulations. 
The combinatorial background is modelled by an exponential function. 
The broad structures are due to \B  and \Lb decays in which 
a final-state particle is either not reconstructed or is 
misidentified as a different hadron, and the 
mass distributions of these backgrounds are derived from simulations.
The \Bs signal yield obtained from the fit is approximately 95,000.  
Candidates in the mass range 5320--5600\mevcc are selected for the
calibration of the SSK algorithm. A fit to the \Bs mass distribution is
performed to extract \sWeights~\cite{Pivk:2004ty}; in this fit the relative fractions of the
background components are fixed 
by integrating the components obtained in the previous fit across the small mass window.
The \sWeights are used to subtract the 
background in the fit to the unbinned distribution 
of the reconstructed \Bs decay time, $t$. 
This procedure for subtracting the background is validated 
with pseudoexperiments and provides unbiased estimates of the calibration parameters.   

The sample is split into three categories --- untagged, mixed and unmixed candidates --- and 
a simultaneous fit to the $t$ distributions of the three subsamples is
performed. 
Untagged candidates are those for which the SSK algorithm cannot make a tagging decision, 
 \ie that contain no tagging tracks passing the $o_1$ selection. 
A \Bs candidate is defined as mixed if the flavour found by the SSK algorithm 
differs from the flavour at decay, determined by the charges 
of the final-state particles; it is defined as unmixed if the flavours are the same. 
The probability density function (PDF) used to fit the $t$ distribution is 
\begin{equation}
P(t)\propto a(t)\left[\Gamma(t') \otimes R(t-t')\right], 
\end{equation}
where $t'$ is the true decay time of the \Bs meson, 
$\Gamma(t')$ is the \Bs decay rate, 
$R(t-t')$ the decay time resolution function, and 
$a(t)$ is the decay time acceptance.

The decay rate of untagged candidates is given by 
\begin{equation}
\Gamma(t') \propto (1- \etag)\, e^{-t' /\tau_s}\, \cosh \left(\frac{\DGs
}{2}t' \right), 
\end{equation}                                              
and that of tagged candidates by 
\begin{equation}
\Gamma(t') \propto \etag \, e^{-t'/\tau_s}\, \left( \cosh\left(\frac{\DGs
}{2}t' \right) + q^{\mathrm{mix}}\, (1- 2\omega)\, \cos (\dms t') \right), \label{gamma_pdf}
\end{equation}
where $q^{\mathrm{mix}}$ is $-1$ or $+1$ for candidates which are mixed or unmixed respectively, and 
$\omega$ is the mistag fraction. 
The average \Bs lifetime, $\tau_s$, the width difference of the \Bs mass eigenstates, \DGs, 
and their mass difference, \dms, are fixed to known values~\cite{LHCb-PAPER-2014-059, LHCb-PAPER-2013-006, PDG2014}.

Each measurement of $t$ is assumed to have a Gaussian uncertainty, $\sigma_t$, which 
is estimated by a kinematic fit of the \Bs decay chain. 
This uncertainty is corrected with a scale factor of 1.37,  
as measured with data from a sample of fake \Bs candidates, which
consist of combinations of a $\Dsm$ candidate and a \pip candidate, both
originating from a primary interaction~\cite{LHCb-PAPER-2013-006}. 
Their decay time distribution is a $\delta$-function at zero 
convolved with the decay time resolution function, $R(t-t^{'})$. 
The latter is described as the sum of three Gaussian functions. 
The functional form of $a(t)$ is modelled with simulated data and its parameters
are determined in the fit to data. 

\begin{figure}[tb]
\begin{center}
\includegraphics[width=.48\columnwidth]{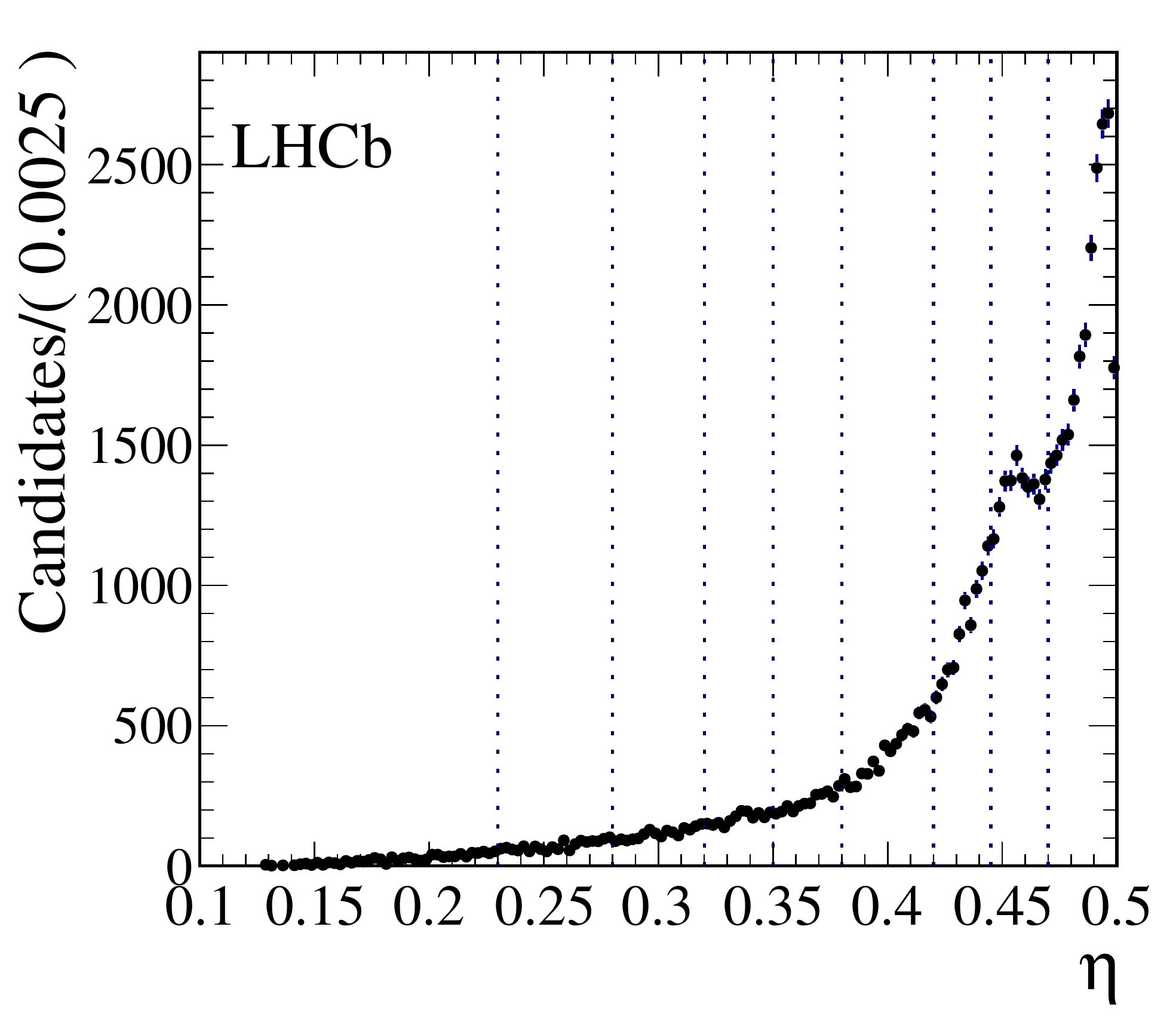}~~
\includegraphics[width=.48\textwidth]{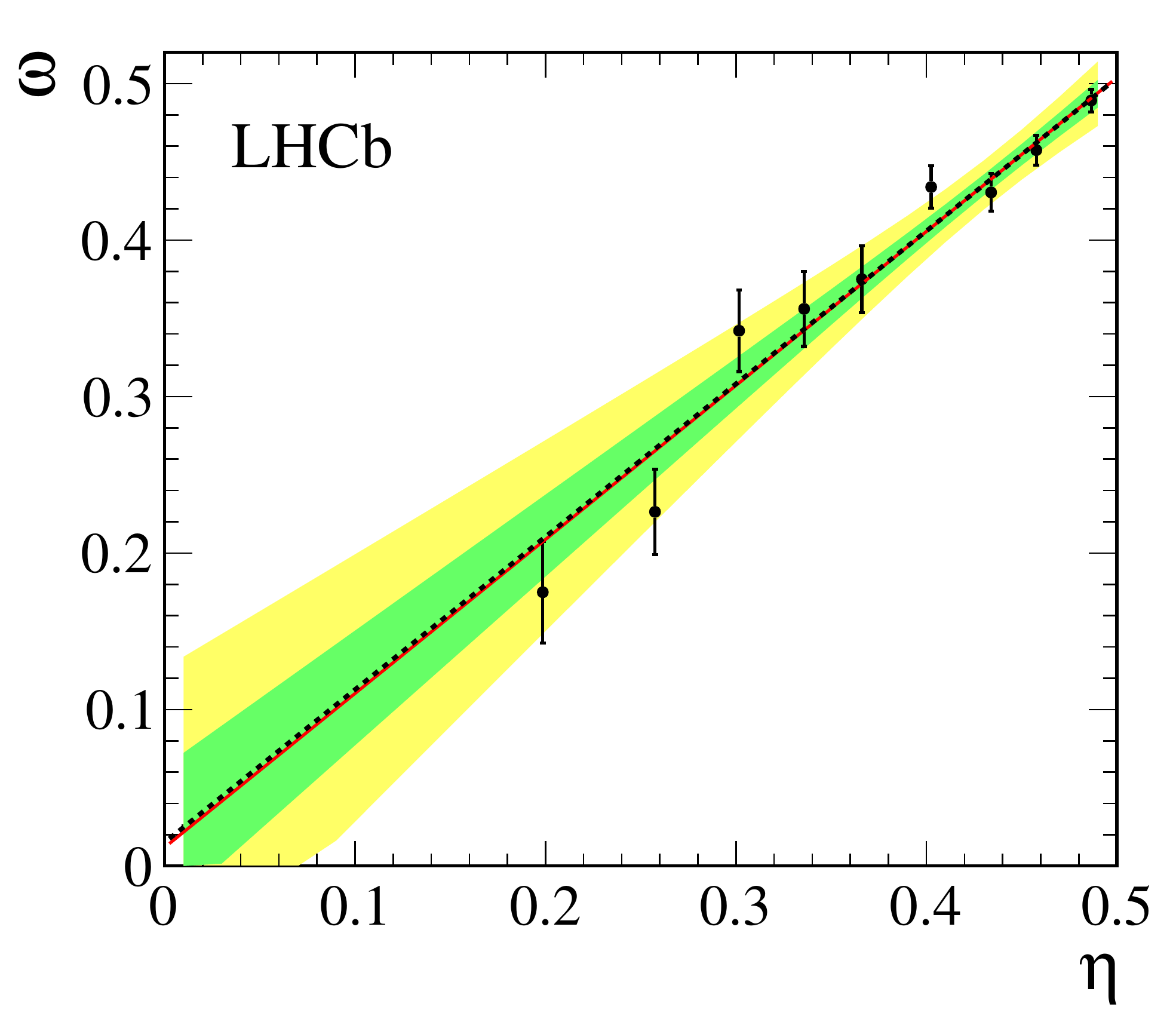}
\caption{(left)~Background-subtracted $\eta$ distribution of  \BsDspi candidates in data; 
the vertical dotted lines show the binning used in the second method of the calibration.  
(right)~Measured average mistag fraction $\omega$ in bins of mistag probability $\eta$ (black points), 
with the result of a linear fit superimposed (solid red line) 
and compared to the calibration obtained from the unbinned fit (dashed black line). 
The linear fit has $\chisqndf=1.3$. The shaded areas correspond to the 68\% and 95\%
confidence level regions of the unbinned fit. }
\label{fig:calib_binned-2011-2012-main}
\end{center}
 \end{figure}

Two methods are used to calibrate the mistag probability. In the first one, 
$\eta$ is an input variable of the fit, and 
 $\omega$ in Eq.~\ref{gamma_pdf} is replaced by the calibration function $\omega(\eta)$,  
 which is assumed to be a first-order polynomial, 
\begin{equation}
\omega(\eta) = p_0 + p_1 (\eta \, - \langle \eta \rangle), \label{omega_calib}
\end{equation}
where $\langle \eta \rangle$ is the average of the $\eta$ 
distribution of signal candidates (Fig.~\ref{fig:calib_binned-2011-2012-main}), fixed to the value 0.4377, 
while $p_0$ and $p_1$ are the calibration parameters to be determined by the fit.  
They are found to be 
\begin{eqnarray}
p_0 - \langle \eta \rangle &=& 0.0052 \pm  0.0044\stat, \nonumber \\ 
p_1 &=& 0.977 \pm  0.070\stat, \nonumber 
\end{eqnarray}
consistent with the expectations of a well-calibrated algorithm,  
 \mbox{$p_0 - \langle \eta \rangle =0$} and \mbox{$p_1=1$}.
The fitted values above are considered as the nominal results of the calibration. 
After  calibration of the mistag probability, the tagging efficiency and tagging power 
measured in \BsDspi decays 
are found to be $\etag = (60.38 \pm 0.16\stat)\%$ and $\effeff=(1.80 \pm 0.19\stat)\%$.  

In the second method, the average mistag fraction $\omega$ is determined by fitting the \Bs decay time distribution split 
into nine bins of mistag probability. Nine pairs $(\langle \eta_{j} \rangle ,\omega_{j})$ are obtained, 
where $\omega_{j}$ is the mistag fraction fitted in the bin $j$, which 
has an average mistag probability $\langle\eta_j\rangle$.
The $(\langle \eta_{j} \rangle ,\omega_{j})$ pairs  are fitted with the calibration function of Eq.~\ref{omega_calib} 
to measure the calibration parameters $p_0$ and $p_1$. 
The calibration parameters obtained, \mbox{$p_0 - \langle \eta \rangle = 0.0050 \pm 0.0045\stat$}
 and \mbox{$p_1=0.983\pm0.072\stat$}, 
are in good agreement with those reported above.
This method also demonstrates the validity of the linear parametrisation (Eq.~\ref{omega_calib}), 
as shown in  Fig.~\ref{fig:calib_binned-2011-2012-main}.  

A summary of the systematic uncertainties on the calibration parameters is given in Table~\ref{tab:syst}.
The dominant systematic uncertainty is due to 
the uncertainty of the scale factor associated with $\sigma_t$. 
The scale factor is varied by $\pm10\%$, the value of its relative uncertainty, and  
the largest change of the calibration parameters due to these variations  is taken as 
 the systematic uncertainty. 
Variations of the functions which describe 
the signal and the background components in the mass fit, and
variations of the fraction of the main peaking background under the signal peak  
due to  $\Bs \rightarrow \Dsm \Kp$ decays, result only in minor changes of the
calibration parameters. The systematic uncertainties associated with these 
variations are assessed by generating pseudoexperiments with a range of
different models and fitting them with the nominal model. 
Systematic uncertainties related to the parametrisation of the acceptance function,  
and to the parameters \DGs, $\tau_s$ and
\dms, are evaluated with the same method; no significant effect
on the calibration parameters is observed. 
The difference between the two 
calibration methods reported in the previous section is assigned as a 
systematic uncertainty.  
Additionally, the calibration parameters are estimated in 
independent samples split according to
different running periods and magnet polarities. No significant differences are observed. 

\begin{table}[tb]
\begin{center}
\caption{Systematic uncertainties of the  parameters
  $p_0$ and $p_1$ obtained in the calibration with \BsDspi decays.} \label{tab:syst}
\begin{tabular}{lcc}
Source & $\sigma_{p_0}$ & $\sigma_{p_1}$ \\
\hline
Decay time resolution                 & 0.0033 & 0.060 \\
Calibration method                    & 0.0002 & 0.006 \\
Signal mass model                     & 0.0001 & 0.002 \\ 
Background mass model                 & 0.0015 & 0.025 \\
$\Bs \rightarrow \Dsm \Kp$ yield      & 0.0001 & 0.008 \\ \hline
Sum in quadrature                     &  0.0036      & 0.066       
\end{tabular}
\end{center}
\end{table}

\section{\boldmath Calibration using $B_{s2}^*(5840)^0 \rightarrow B^+K^-$  decays}
\label{Bsstar}
In \mbox{\BstwoBK} decays, 
the \Bp candidates are reconstructed in four exclusive final states,  
$\Bp \to \jpsi (\to \mup \mun) \Kp$, $\Bp \to \Dzb (\to \Kp\pim) \pip$, $\Bp
\to \Dzb(\to \Kp\pim) \pip \pim \pip$ and
\mbox{$\Bp\to \Dzb(\to \Kp \pim \pip \pim)\pip$}.
The \Bp candidate selection follows the same strategy as in 
Ref.~\cite{LHCb-PAPER-2012-030},
retaining only those candidates with a \Bp mass in the range 5230--5320\mevcc. 
The \Bp candidate is then combined with 
a \Km candidate to form a common  
vertex.  Combinatorial background is reduced by requiring the \Bp and \Km candidates to have
a minimum \pt of 2000\mevc and 250\mevc respectively, 
and to be compatible with coming from the PV. The kaon candidate must have  good particle 
identification and a minimum momentum of 5000\mevc. 
A good-quality vertex fit of the $\Bp\Km$ combination is required.  
 In order to improve the mass resolution, the invariant mass of the system, $m_{\Bp\Km}$,
 is computed constraining the masses of the \jpsi (or \Dz) and \Bp candidates to their
  world average values~\cite{PDG2014} and constraining the vector momenta of 
\Bp and \Km candidates to point to the associated primary vertex. 
Finally, the \Bp\Km system is required to have a minimum transverse momentum
of 2500\mevc. 

\begin{figure}[tb]
  \begin{center}
    \includegraphics[width=0.80\linewidth]{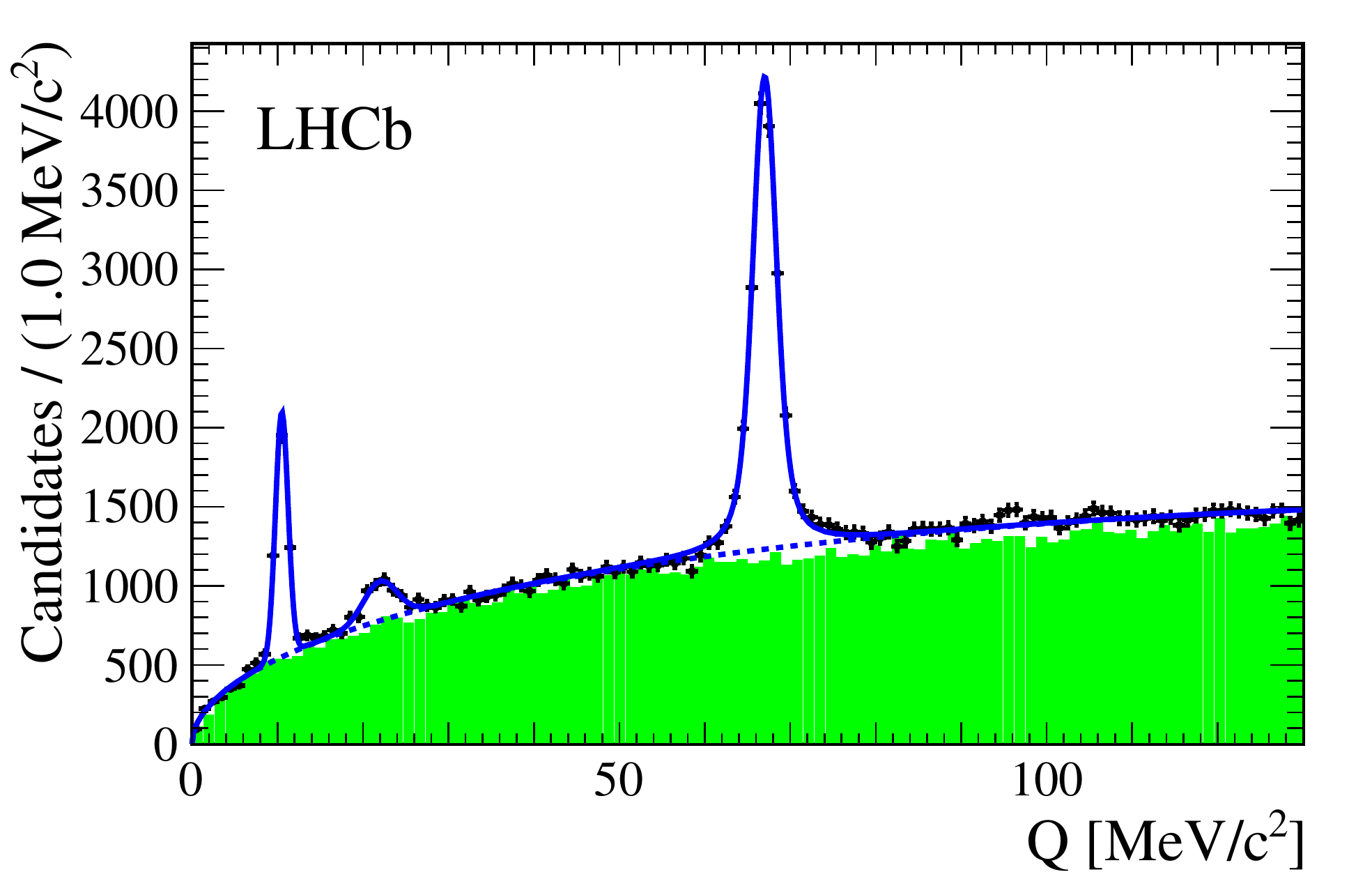}
  \end{center}
  \caption{Distribution of the mass difference, $Q$, of
selected $\Bu\Km$ candidates, summing over
four \Bu decay modes (black points), and the function fitted to
these data (solid blue line).
From left to right, the three peaks are identified as being $\B_{\squark1}(5830)^0\to \B^{*+}\Km$, 
$\Bstwo \to \B^{*+}\Km$, and \BstwoBK. 
Same charge combinations $\Bpm\Kpm$ in data are superimposed (solid histogram) and contain no structure.
}
  \label{fig:deltam_all}
\end{figure}
The mass difference, $Q \equiv m_{\Bp\Km}-M_{\Bp}-M_{\Km}$,
where $M_{\Bp}$ and $M_{\Km}$ are the nominal masses of the \Bp and  
\Km mesons, is shown in Fig.\,\ref{fig:deltam_all} for the selected 
$\Bp\Km$ candidates, summed over all the \Bp decay modes.
The spectrum is consistent with that seen in Ref.\,\cite{LHCb-PAPER-2012-030} 
and contains three narrow peaks at $Q$-values of approximately 11, 22 and 67\mevcc, 
which are interpreted as \mbox{$\B_{\squark1}(5830)^0 \to \B^{*+}(\to \Bu \g)\Km$}, 
\mbox{$\Bstwo \to \B^{*+}(\to \Bu \g)\Km$} and 
\mbox{\BstwoBK}, respectively. 
The first two peaks are shifted down by \mbox{$M_{\B^{*+}}-M_{\Bu}=45.0 \pm 0.4$\mevcc}  from to their 
nominal $Q$-values due to the unreconstructed photons in the $\B^{*+}$ decays.

The yields of the three peaks are obtained through a
fit of the $Q$ distribution in the range shown.
Both the $\B_{\squark1}(5830)^0\to \B^{*+}\Km$ and the $\Bstwo \to \B^{*+}\Km$ signals 
are described by Gaussian functions. 
The \BstwoBK signal is parametrised as a
relativistic Breit-Wigner function convolved with a Gaussian function 
to account for the detector resolution. 
This resolution is fixed to the value determined in the simulation ($\simeq1$\mevcc).
The background is modelled by the function
$f(Q)=Q^{\alpha} e^{\beta  Q}$, where $\alpha$ and $\beta$ are free parameters.
The yields of the three peaks are found to be approximately 
2,900, 1,200 and 12,700, respectively. 
The mass and width parameters are in agreement with those obtained in 
Ref.~\cite{LHCb-PAPER-2012-030}. 
Only the third peak, corresponding to the 
fully reconstructed \Bstwo meson, is used in the calibration of the 
mistag probability.

\begin{figure}[tb]
\begin{center}
\includegraphics[width=0.48\linewidth]{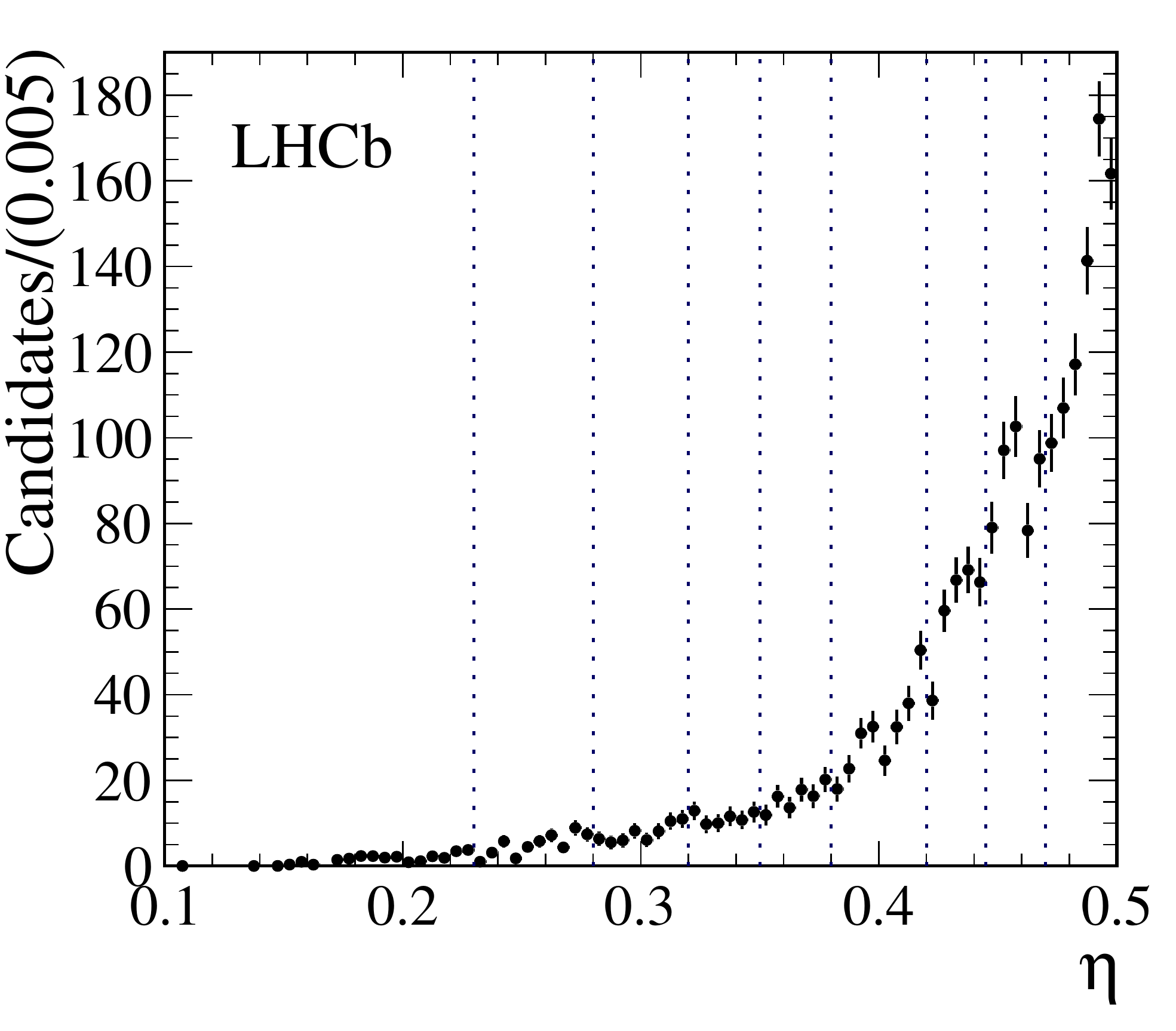} ~~
\includegraphics[width=0.48\linewidth]{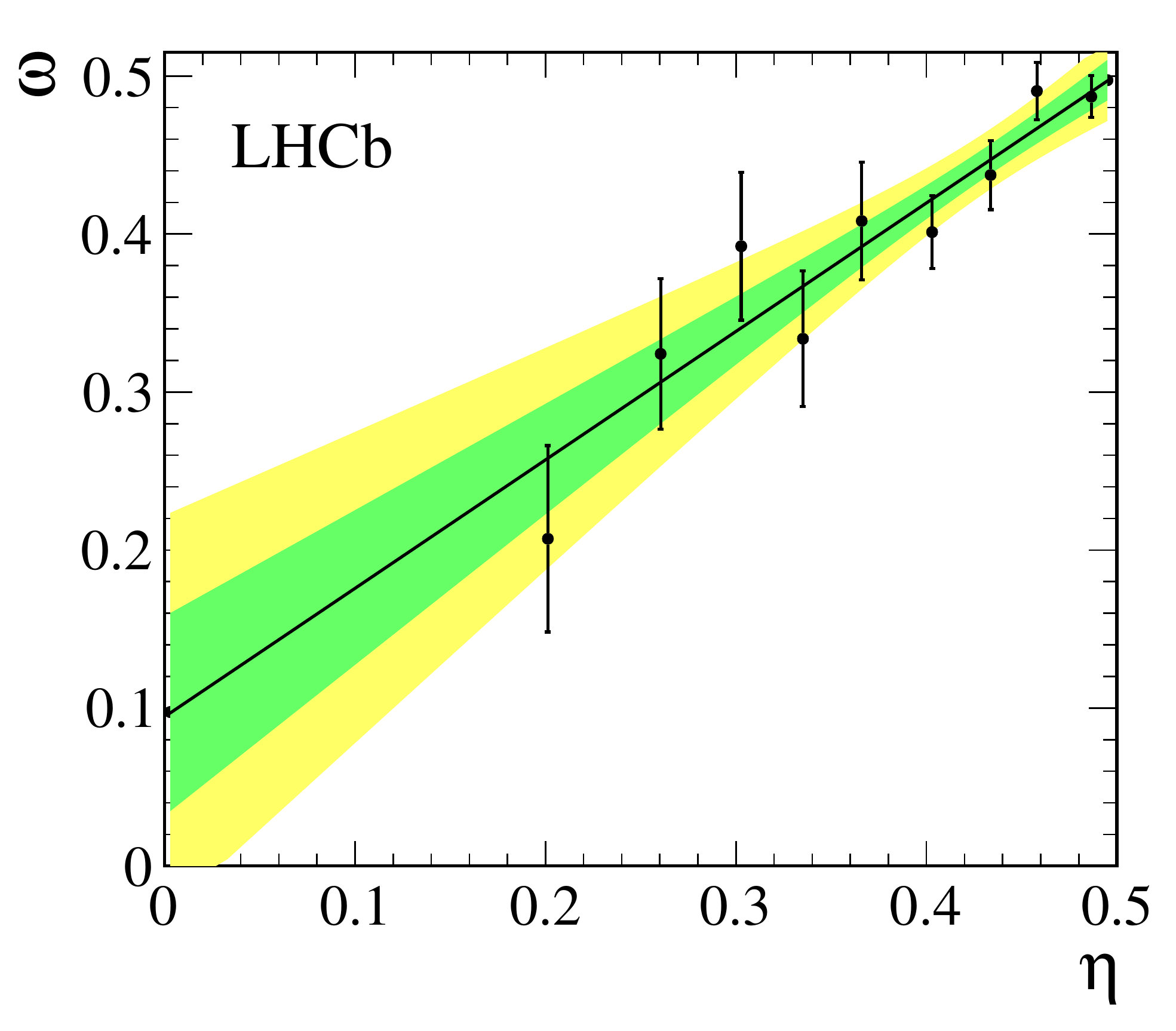} 
\caption{
(left)~Background-subtracted $\eta$ distribution of \BstwoBK candidates in data;
 the vertical dotted lines show the binning used in the calibration. 
(right)~Measured average mistag fraction $\omega$ in bins of mistag probability $\eta$ (black points), 
with the result of a linear fit superimposed (solid black line). 
The fit has $\chisqndf=0.8$. 
The shaded areas correspond to the 68\% and 95\%
confidence level regions of the fit. 
}
\label{fig:results_sw_1}
\end{center}
\end{figure}

Since the \Bstwo meson is flavour-tagged by the charges of the final-state 
particles of its decay, the mistag fraction 
can be determined by comparing the tagging decision
of the SSK algorithm with the known \Bstwo flavour. 
From the fit of the $Q$ distribution, {\it sWeights}
are obtained and used to statistically disentangle the signal from the 
combinatorial background.
The fit is performed separately on the $Q$ distributions of correctly and
incorrectly tagged candidates, to allow for different background fractions in the two categories. 
In these fits the mass parameters are fixed to the values obtained in the fit to all candidates.
In Fig.~\ref{fig:results_sw_1} the $\eta$ distribution of signal candidates 
and the mistag fraction $\omega$ in bins of $\eta$ are shown.
Each bin of $\eta$ has an average predicted mistag $\langle\eta\rangle$.
The $(\langle \eta \rangle ,\omega)$ pairs are fitted with 
the calibration function of Eq.~\ref{omega_calib} to determine 
the calibration parameters.

The calibration parameters depend on the kinematics of the reconstructed \B meson, 
and in particular on its transverse momentum.
In order to test whether the calibrations are consistent between the two samples, 
the \Bstwo \pt spectrum must be reweighted to match that of 
the \Bs candidates seen in \BsDspi decays. 
This is done for each of the four \Bu decay modes separately. 
Due to the requirement of a higher minimum \pt of
 the \Bstwo candidates, 2.5\gevc, 
compared to 2.0\gevc for the \Bs candidates, a 1\% 
  difference in the mean value of the \pt spectra remains.  
This is covered by the systematic uncertainties discussed 
in Section~\ref{sec:systematics}, which account for differences in the mean transverse momenta
 of \B mesons of up to 30\%.
The calibration parameters obtained from the full sample of weighted
\Bstwo decays are 
\begin{eqnarray}
p_0-\langle \eta\rangle &=& 0.012 \pm 0.008\stat, \nonumber \\
p_1 &=& 0.813 \pm 0.123\stat, \nonumber 
\end{eqnarray}
where $\langle \eta \rangle$ is fixed to the value 0.441.
They are consistent within statistical uncertainties with the 
calibration parameters obtained with \BsDspi decays.

 \begin{table}[tb]
\caption{Systematic uncertainties of the  parameters
  $p_0$ and $p_1$ obtained in the calibration with \BstwoBK decays.}
\begin{center}\begin{tabular}{lcc}
 Source & $\sigma_{p0} $ &$\sigma_{p1} $ \\
\hline
 Signal model & 0.0063 & 0.012\\
 Background model & 0.0008 & 0.054\\
\kaon from \Bstwo \pt selection & 0.0028 & 0.039\\
\kaon from \Bstwo particle identification & 0.0025 & 0.015\\
\hline
 Sum in quadrature & 0.0074 & 0.069\\
\end{tabular}\end{center}
\label{tab:syst_BDSt}
\end{table}
The systematic uncertainties of the calibration parameters are determined 
by repeating the calibration under different conditions.
In each case the fit to the $Q$  distribution is repeated and the 
{\it sWeights} are calculated. 
A summary of all of the systematic uncertainties is given in Table~\ref{tab:syst_BDSt}.
To test for potential differences in the signal model for correctly and incorrectly tagged
 candidates, the  fit to the $Q$ distribution is repeated for both subsets of \Bstwo candidates
without fixing the mass parameters to the values obtained in the fit to all candidates.
The background fit model is tested by fitting the $Q$ distribution of correctly and
incorrectly tagged candidates
with the default background model replaced by
 a second-order polynomial, and with the fit range limited to 
 $40 < Q < 100\mevcc$.  
The mass resolution for \Bstwo is varied by $\pm 10\%$ to account for 
differences in resolution between data and simulation.
Potential biases due to the \Bstwo  signal selection are studied by varying
the requirements on the \pt or on the particle identification 
probability of the kaon produced in the \Bstwo decay and repeating the full calibration 
procedure.
To test the background subtraction procedure, an alternative method 
of performing the calibration is used. The sample of tagged candidates is divided
into bins of $\eta$, and, in each bin, the $Q$ distributions of correctly and incorrectly
tagged candidates are fitted separately. 
The measured signal yields of the \Bstwo peak are used to calculate the 
mistag fraction $\omega$ which is plotted against the average $\eta$ of each bin.
The calibration parameters obtained  are in agreement  within statistical uncertainties 
with those determined from the default method.

The variation of the calibration parameters with  data-taking 
conditions is checked by repeating the calibration procedure after splitting the 
candidate sample according to the data-taking period 
and magnet polarity. No significant variation is observed.
The calibration is also repeated separately on each of the four \Bp decay 
modes, after weighting the transverse momentum spectra. 
The parameters obtained  agree within statistical uncertainties.

\section{Portability to different decay channels}
\label{sec:systematics}

The tagging calibration parameters will in general depend on the kinematics of the reconstructed \B candidate
and on the properties of the event. The largest dependences are found to be on the \pt of the \B
candidate and on the track multiplicity of the event.   
The calibration parameters measured in \BsDspi and \BstwoBK decays can
thus be used in decays which have similar distributions in these variables. 
This is not necessarily the case for all \Bs decay modes, due to different trigger and selection requirements. 
Three representative \Bs decay modes have been studied: 
$\decay{\Bs}{\jpsi\phi}$, $\Bs \rightarrow \Dsp \Dsm$ and $\Bs
\rightarrow \phi \phi$. 
The sample of \BsDspi candidates is weighted to match the \B meson \pt 
and event track multiplicity distributions of each of the three other decay modes in turn,
 with the weighting done for each variable separately.
For each of the weighted
samples, $p_0$ and $p_1$ are measured  and compared to those of the unweighted sample. 
For each calibration parameter, a systematic uncertainty due to decay mode dependence is assigned, 
equal to half of the largest difference seen between the unweighted and weighted \BsDspi samples.
The systematic uncertainties obtained are listed in Table~\ref{tab:syst_bis}. 
The dominant effect is due to the weighting to match the \pt distribution.

\begin{table}[tb]
\begin{center}
\caption{Systematic uncertainties of the parameters
  $p_0$ and $p_1$ related to the portability of the calibration
   to different decay modes.} \label{tab:syst_bis}
\begin{tabular}{lcc}
Source & $\sigma_{p_0}$ & $\sigma_{p_1}$ \\
\hline
Weighting in \pt & 0.0011 & 0.030 \\
Weighting in track multiplicity & 0.0006 & 0.006 \\ \hline
Sum in quadrature &  0.0012 &   0.031\\
\end{tabular}
\end{center}
\end{table}

\section{Flavour-tagging asymmetry}
\label{sec:asymmetry}

The calibration parameters depend on the 
initial flavour of the \Bs meson, due to the different interaction cross-sections of \Kp and
\Km with matter. Therefore, additional calibration parameters, $\Delta p_0$
and $\Delta p_1$, are introduced to take this flavour dependence into account.
The mistag fraction of mesons produced with initial flavour \Bs (accompanied by a \Kp) 
and mesons produced with initial flavour \Bsb (accompanied by a \Km) are given by
\begin{eqnarray}
\omega(\eta) &=& p_0 + \frac{\Delta p_0}{2} + \left( p_1 + \frac{\Delta
  p_1}{2} \right)(\eta \, - \langle \eta \rangle) {\mathrm{~and}} \label{B0calib}\\
\overline{\omega}(\eta) &=& p_0 -  \frac{\Delta p_0}{2} + \left( p_1 - \frac{\Delta
  p_1}{2} \right) (\eta \, - \langle \eta \rangle), \label{B0barcalib}
\end{eqnarray}
respectively. The statistical power of the 
\BsDspi\ data sample is not sufficient to determine these additional  
parameters, so  they are studied with 
\mbox{$\Dsm \rightarrow \phi(\to \Kp\Km)\pim$} decays. 
The \Dsm mesons produced in the primary interaction 
are also accompanied by charged kaons
produced in the \cquark quark hadronisation. 
The SSK algorithm can tag the initial flavour of the \Dsm candidate, with 
a tagging decision opposite to the case of \Bs mesons. 
The \Dsm meson is charged and does not oscillate, so its initial flavour can be determined 
from the charge of the decay products. This can then be compared to the SSK tagging decision, 
and a calibration can be performed with the same method used with \BstwoBK decays.  
The $\Delta p_0$ and $\Delta p_1$ parameters can be determined by 
the difference in the calibration parameters obtained with $\Dsm$ and $\Dsp$ decays.

\begin{figure}[tb]
 \centering
\includegraphics[width=.5\textwidth]{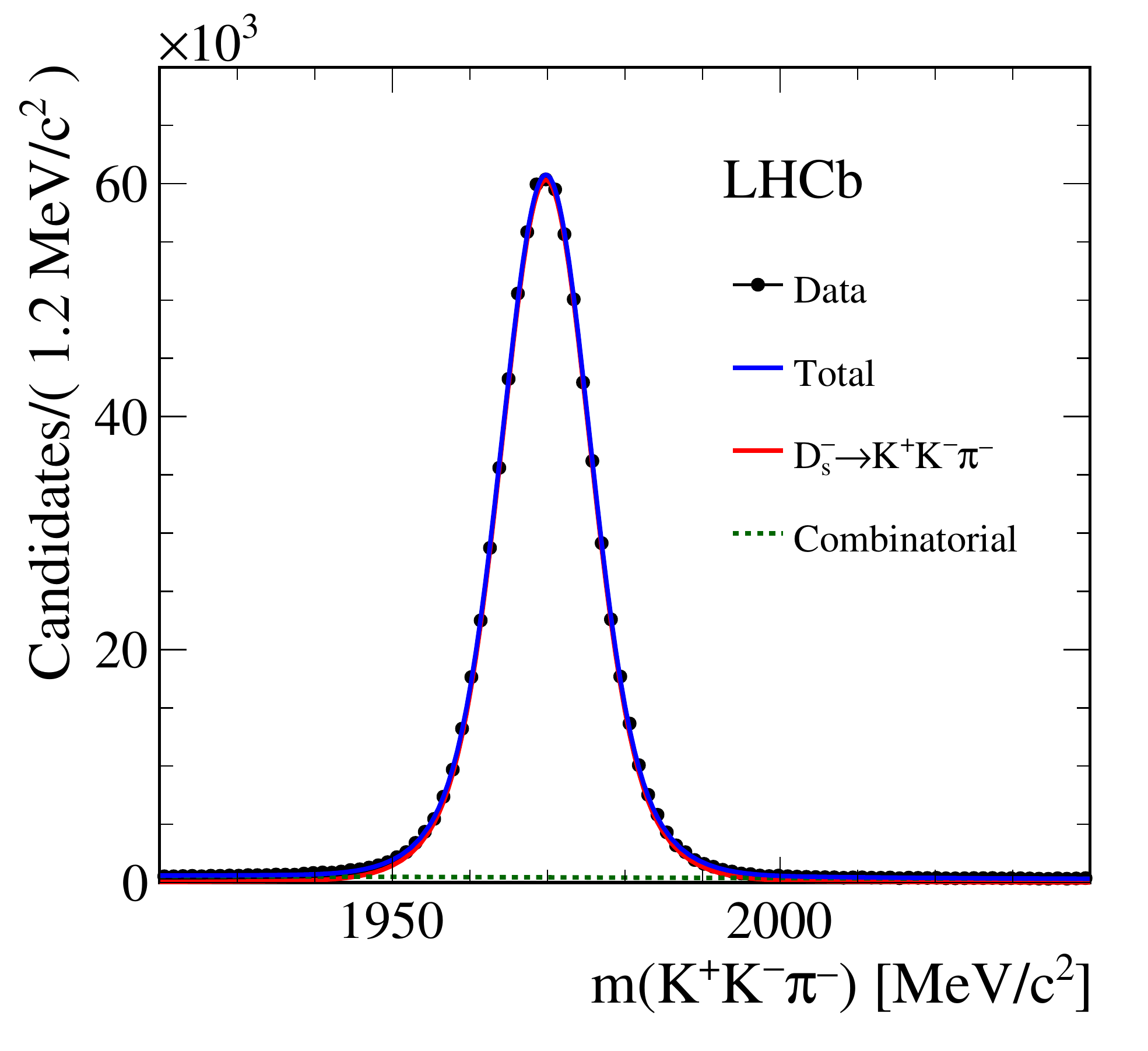}
\caption{Mass distribution of \mbox{$\Dsm \rightarrow \phi(\to \Kp\Km)\pim$}  candidates with fit
projections overlaid.  
Data points (black markers) correspond to the \Dsm candidates selected in the 3\invfb data sample. 
The total fit function and its components are overlaid (see legend). 
}\label{fig:Ds_mass_fit}
\end{figure}
A high-purity sample of \mbox{$\Dsm \rightarrow \phi(\to \Kp\Km)\pim$} candidates 
is selected  in a sample corresponding 
to  3\invfb of data taken at centre-of-mass energies of 7
and 8\tev by applying the following criteria. The momenta of the final-state 
 particles must be larger than 2\gevc and their transverse momenta larger than 250\mevc. 
The tracks must be significantly displaced from the primary
vertex. Their associated particle type information is required to be consistent with a
kaon or a pion, as appropriate. 
The $\Kp\Km$ invariant mass must be within 7\mevcc of the known
$\phi$ mass. The $\phi$ and the \Dsm reconstructed vertices must be
of good quality. 
The momentum vector of the \Dsm candidate must be consistent 
with the displacement vector between the primary vertex and the \Dsm decay vertex. 
Only candidates with a reconstructed \Dsm mass in the
range 1920--2040\mevcc are considered.  
The resulting \Dsm mass distribution is fitted 
by a sum of two Gaussian functions with a common mean to describe the signal component, 
 and an exponential function for the combinatorial
background (Fig.~\ref{fig:Ds_mass_fit}). In total about 784,000 signal candidates are reconstructed with a 
background fraction below 5\%. From the mass fit, \sWeights are calculated to 
subtract the background in the $\eta$ distributions of correctly and incorrectly tagged \Dsm candidates. 
 Differences between the \Dsm and the \Bs
kinematics are accounted for by weighting the \Dsm candidates to
match the $B^0_s$ transverse momentum distribution measured with \BsDspi decays. 
The average mistag probability in Eq.~\ref{eq:parameter_asymmetries} is fixed to 
the value found for  \BsDspi decays, 0.4377. The parameters related to the flavour-tagging 
asymmetries are found to be                                             
\begin{eqnarray}  
\label{eq:parameter_asymmetries}                                                                                                          
\Delta p_0 &=& -0.0163 \pm 0.0022\stat \pm 0.0030\syst,
\nonumber \\                                                                                                                       
\Delta p_1 &=& -0.031  \pm 0.025\stat  \pm 0.045\syst,
\nonumber \\                                                                                                                        
\Delta \etag &=& (0.17 \pm 0.11\stat \pm 0.68\syst)\%,                                                    
\end{eqnarray}                                                                                                                                                       
where $\Delta \etag\equiv \etag(\Dsm) -
\etag(\Dsp) = \etag(\Bs) -
\etag(\Bsb)$. 

A systematic uncertainty is computed by taking the maximum 
of the differences seen when comparing these calibration parameters 
and those obtained by weighting the transverse momentum distribution of the 
\Dsm candidates to match the following \Bs decay modes:
\mbox{$\decay{\Bs}{\jpsi\phi}$}, \mbox{$\Bs \rightarrow \phi \phi$}
\mbox{$\Bs \rightarrow \Dsp \Dsm$}. 
These uncertainties are 0.0030 and 0.040 for $\Delta p_0$ and $\Delta p_1$ respectively, 
and 0.66\% for $\Delta \etag$.
The same procedure is applied
to assess the systematic uncertainty associated with the different
track multiplicity distribution between \Ds and \Bs decays 
(0.0002 and 0.020 for $\Delta p_0$ and $\Delta p_1$ respectively, and 0.15\% for $\Delta \etag$). 
The systematic uncertainty in Eq.~\ref{eq:parameter_asymmetries} is the sum in quadrature of 
these two sources of uncertainties. 

While the shift of the slope parameter $\Delta p_1$ is compatible with 
zero, there is a significant overall shift, $\Delta p_0$, of about 1.6\%
towards higher mistag rates for \Bsb particles. 
This can be explained by the higher interaction rate 
in matter of \Km particles compared to \Kp particles. These values are 
consistent with results obtained in simulated samples of \BsDspi 
 and \mbox{$\decay{\Bs}{\jpsi\phi}$} decays.
 
The \Bstwo decays can also be used to measure                                                             
the values of $\Delta p_0$, $\Delta p_1$ and $\Delta \etag$.                                                                    
The \Bstwo candidates are split into two samples according to the 
final-state charges, $\Bu \Km$ and $\Bub\Kp$, 
and the calibration described in Sect.~\ref{Bsstar} is performed in the two samples.                                                                 
The differences of the calibration parameters between \Bstwo
and {{\ensuremath{\Bb_{\squark2}^{*}(5840)^0}}\xspace} are $\Delta p_0=-0.01 \pm 0.02\stat$ and 
$\Delta p_1=-0.4 \pm 0.2\stat$, and $\Delta \etag =(-1.4 \pm 1.3\stat)$\%.
They are compatible with the shifts measured in the prompt \Dsm meson sample.

\section{Calibration summary}
\label{sec:calibration_summary}

The final calibration parameters are
computed as the weighted average of the 
results obtained in \BsDspi and \BstwoBK decays, 
 fixing $\langle \eta \rangle = 0.4377$ 
and considering the systematic uncertainties
reported in Tables~\ref{tab:syst} and \ref{tab:syst_BDSt} to be uncorrelated. 
The uncertainties relating to the portability of the calibrations  
to different \Bs decays
as reported in Table~\ref{tab:syst_bis} are considered to be fully correlated. 
For the flavour-tagging asymmetries, only the results measured in \Dsm decays are considered.  
The final values are 
\begin{eqnarray}
\langle \eta \rangle &=& 0.4377, \nonumber \\
p_0- \langle \eta \rangle &=& 0.0070 \pm 0.0039\stat \pm 0.0035\syst, \nonumber \\ 
p_1 &=& 0.925 \pm 0.061\stat \pm 0.059\syst, \nonumber \\ 
\Delta p_0 &=& -0.0163 \pm 0.0022\stat \pm 0.0030\syst, \nonumber \\
\Delta p_1 &=& -0.031  \pm 0.025\stat  \pm 0.045\syst, \nonumber \\
\Delta \etag &=& (0.17 \pm 0.11\stat \pm 0.68\syst)\%.   \nonumber
\end{eqnarray}

\section{ Possible application to OS kaons}
\label{sec:note}
The two-step neural-network approach 
of the SSK tagging algorithm presented here  
is a promising method  for improving 
any tagging algorithm which needs to combine  information from
multiple tagging tracks.
 A natural candidate  for the application 
of this method is the OS kaon tagging algorithm, which searches for kaons from
$b \rightarrow c \rightarrow s$ transitions of 
the OS \bquark hadron.  
The current implementation of the OS kaon algorithm selects tracks
with large impact parameters with respect to the primary vertex
associated with the signal \B meson~\cite{LHCb-PAPER-2011-027}.
This selection gives a tagging efficiency of about 15\%. 
A preliminary implementation of a neural-network-based algorithm
shows that loosening the impact parameter requirements for the 
track candidates and using the new approach increases the
tagging efficiency to about 70\% and significantly improves the effective
tagging efficiency of \Bp and \Bd mesons.  
However, the inclusion of kaons with smaller impact parameters results in up to 10\% of the signal fragmentation tracks being assigned as OS kaon candidates. As the correlation of
signal fragmentation kaons with the signal \B flavour is different for \Bu, \Bd and \Bs mesons, this
contamination of SS kaon tracks introduces a dependence of 
the  calibration parameters on the \B meson species, and the gain in tagging
performance observed in \Bu and \Bd is not reproduced in \Bs mesons.

\section{Conclusion}
\label{sec:conclusion}
 A new algorithm for the determination of the 
flavour of \Bs mesons at production has been presented. The algorithm is
based on two neural networks, the first trained to select charged kaons produced in
association with the \Bs meson, and the second to combine the kaon charges to
assign the \Bs flavour, and to estimate the probability of an incorrect flavour assignment. 
The algorithm is calibrated with data corresponding to an integrated luminosity of 3\invfb
 collected by the LHCb experiment in proton-proton collisions at 7 and 8\tev centre-of-mass energies. 
The calibration is performed in two ways: by resolving the \mbox{\Bs-\Bsb}
flavour oscillations in \BsDspi decays, and, for the first time, by analysing
flavour-specific \BstwoBK strong decays. 
 
The tagging power of the new algorithm as measured in \mbox{\BsDspi}  decays is 
\mbox{$(1.80 \pm 0.19\stat \pm 0.18\syst )\%$}, 
 a significant improvement over the  tagging power of 1.2\% 
 of the previous implementation used at the LHCb experiment.
This new algorithm represents important progress for many analyses aiming to make 
high-precision measurements of \Bs-\Bsb mixing and \CP asymmetries of \Bs decays.  
Its performance  has been demonstrated  
in several recent measurements  by  
the LHCb collaboration~\cite{LHCb-PAPER-2014-059,LHCb-PAPER-2014-019,
LHCb-PAPER-2014-026,LHCb-PAPER-2014-064,LHCb-PAPER-2014-051}.

\section*{Acknowledgements}

\noindent We express our gratitude to our colleagues in the CERN
accelerator departments for the excellent performance of the LHC. We
thank the technical and administrative staff at the LHCb
institutes. We acknowledge support from CERN and from the national
agencies: CAPES, CNPq, FAPERJ and FINEP (Brazil); NSFC (China);
CNRS/IN2P3 (France); BMBF, DFG and MPG (Germany); INFN (Italy); 
FOM and NWO (The Netherlands); MNiSW and NCN (Poland); MEN/IFA (Romania); 
MinES and FANO (Russia); MinECo (Spain); SNSF and SER (Switzerland); 
NASU (Ukraine); STFC (United Kingdom); NSF (USA).
We acknowledge the computing resources that are provided by CERN, IN2P3 (France), KIT and DESY (Germany), INFN (Italy), SURF (The Netherlands), PIC (Spain), GridPP (United Kingdom), RRCKI and Yandex LLC (Russia), CSCS (Switzerland), IFIN-HH (Romania), CBPF (Brazil), PL-GRID (Poland) and OSC (USA). We are indebted to the communities behind the multiple open 
source software packages on which we depend.
Individual groups or members have received support from AvH Foundation (Germany),
EPLANET, Marie Sk\l{}odowska-Curie Actions and ERC (European Union), 
Conseil G\'{e}n\'{e}ral de Haute-Savoie, Labex ENIGMASS and OCEVU, 
R\'{e}gion Auvergne (France), RFBR and Yandex LLC (Russia), GVA, XuntaGal and GENCAT (Spain), Herchel Smith Fund, The Royal Society, Royal Commission for the Exhibition of 1851 and the Leverhulme Trust (United Kingdom).

\addcontentsline{toc}{section}{References}
\setboolean{inbibliography}{true}
\bibliographystyle{LHCb}
\bibliography{main,LHCb-PAPER,LHCb-DP}

\newpage

%%%%%%%%%%%%%%%%%%%%%%%%%%%%%%%%%%%%%%%%%%
\centerline{\large\bf LHCb collaboration}
\begin{flushleft}
\small
R.~Aaij$^{39}$, 
C.~Abell\'{a}n~Beteta$^{41}$, 
B.~Adeva$^{38}$, 
M.~Adinolfi$^{47}$, 
A.~Affolder$^{53}$, 
Z.~Ajaltouni$^{5}$, 
S.~Akar$^{6}$, 
J.~Albrecht$^{10}$, 
F.~Alessio$^{39}$, 
M.~Alexander$^{52}$
% ...

{\footnotesize \it
% ...
$ ^{5}$Clermont Universit\'{e}, Universit\'{e} Blaise Pascal, CNRS/IN2P3, LPC, Clermont-Ferrand, France\\
$ ^{6}$CPPM, Aix-Marseille Universit\'{e}, CNRS/IN2P3, Marseille, France\\
% ...
$ ^{10}$Fakult\"{a}t Physik, Technische Universit\"{a}t Dortmund, Dortmund, Germany\\
% ...
$ ^{38}$Universidad de Santiago de Compostela, Santiago de Compostela, Spain\\
$ ^{39}$European Organization for Nuclear Research (CERN), Geneva, Switzerland\\
% ...
$ ^{41}$Physik-Institut, Universit\"{a}t Z\"{u}rich, Z\"{u}rich, Switzerland\\
% ...
$ ^{47}$H.H. Wills Physics Laboratory, University of Bristol, Bristol, United Kingdom\\
% ...
$ ^{52}$School of Physics and Astronomy, University of Glasgow, Glasgow, United Kingdom\\
$ ^{53}$Oliver Lodge Laboratory, University of Liverpool, Liverpool, United Kingdom\\
% ...
}
\end{flushleft}
%%%%%%%%%%%%%%%%%%%%%%%%%%%%%%%%%%%%%%%%%%

\end{document}